\documentclass[12pt,letterpaper]{article}

\usepackage{graphicx,array}
\usepackage{url}
\usepackage{color}
\usepackage{latexsym}
\usepackage{amsthm}
\usepackage{amsmath}
\usepackage{amssymb}
\usepackage{amsfonts}
\usepackage[numbers,sort&compress]{natbib}
\usepackage{bm}
\usepackage{slashed}
\usepackage{mathrsfs}
\usepackage{enumerate}
\usepackage{tikz}
\usepackage{siunitx}
\usepackage{mdframed}
\usepackage{setspace}  
\usepackage{esvect}
\usepackage{multirow}

\usepackage[utf8x]{inputenc}%
\usepackage{tcolorbox}%

\usepackage{hyperref} 
\hypersetup{
    colorlinks=true,       
    linkcolor=red,          
    citecolor=blue,        
    filecolor=magenta,      
    urlcolor=blue           
}
\usepackage[all]{hypcap} 

\usepackage{natbib}
\newcommand{\nc}{\newcommand}  
\newcommand{\mc}{\mathcal}
\newcommand{\uu}{\;\!}

\nc{\beq}{\begin{equation}}
\nc{\eeq}{\end{equation}}
\nc{\beqa}{\begin{eqnarray}}  
\nc{\eeqa}{\end{eqnarray}}  
\nc{\bit}{\begin{itemize}}  
\nc{\eit}{\end{itemize}}  

\def\GeV{\mathrm{GeV}}     

\newcommand{\eg}{{\it e.g.}}
\newcommand{\ie}{{\it i.e.}}
\newcommand{\Mpl}{M_{\rm pl}}

\usepackage{floatrow}
\newfloatcommand{capbtabbox}{table}[][\FBwidth]

\usepackage{blindtext}

\newcommand{\lc}{\lambda_{\phi S}}

\newcommand{\lphi}{\lambda_\phi}

\newcommand{\ls}{\lambda_S}
\newcommand{\ms}{m_{S,0}}

\newcommand{\epsc}{\epsilon_c}
\newcommand{\epsn}{\epsilon_n}

\newcommand{\rb}{\overline{r}}

\newcommand{\Qmax}{Q_\text{max}}
\newcommand{\Qmin}{Q_\text{min}}
\newcommand{\Sbar}{S^\dagger}
\newcommand{\Trho}{T_\rho}
\newcommand{\TFD}{T_{F}}

\title{ 
 {\bf Origin of nontopological soliton dark matter:} \\
 {\bf \Large solitosynthesis or phase transition}
\author{\large Yang Bai$^{\,a}$, Sida Lu$^{\,b,c}$, and Nicholas Orlofsky$^{\,d}$}
\date{\small \it 
$^a$Department of Physics, University of Wisconsin-Madison, Madison, WI 53706, USA\\
$^b$School of Physics and Astronomy, Tel Aviv University, Tel Aviv 69978, Israel \\
$^c$Institute for Advanced Study, The Hong Kong University of Science and Technology, \\Clear Water Bay, Kowloon, Hong Kong S.A.R., P. R. China \\
$^d$Department of Physics, Carleton University, Ottawa, ON K1S 5B6, Canada \\
}
}

\begin{document}

\maketitle

\setlength{\parskip}{0.2ex}

\begin{abstract}	
This work demonstrates that nontopological solitons with large global charges and masses, even above the Planck scale, can form in the early universe and dominate the dark matter abundance. In solitosynthesis, solitons prefer to grow as large as possible under equilibrium dynamics when an initial global charge asymmetry is present. Their abundance is set by when soliton formation via particle fusion freezes out, and their charges are set by the time it takes to accumulate free particles. This work improves the estimation of both quantities, and in particular shows that much larger-charged solitons form than previously thought. The results are estimated analytically and validated numerically by solving the coupled Boltzmann equations. Without solitosynthesis, phase transitions can still form solitons from particles left inside false-vacuum pockets and determine their present-day abundance and properties. Even with zero charge asymmetry, solitons formed in this way can have very large charges on account of statistical fluctuations in the numbers of (anti)particles inside each pocket.\end{abstract}

\thispagestyle{empty}  
\newpage  
  
\setcounter{page}{1}  

\begingroup
\hypersetup{linkcolor=black,linktocpage}
\tableofcontents
\endgroup

\newpage

\section{Introduction}
\label{sec:intro}
Nontopological solitons (NTSs) are fascinating macroscopic states that may exist in theories containing scalar or fermion fields with a conserved global symmetry and nonlinear interactions.
Once its global charge is above a minimum charge, an NTS could be stable at the quantum level if it is energetically forbidden to decay into states with smaller charges.
Because of their longevity at the cosmological scale, NTSs could account for all or part of the dark matter in our Universe.  
Their properties and detection methods are also different from ordinary dark matter searches for point-like particles, which make them interesting objects to study.
Historically, NTSs were proposed by G. Rosen~\cite{Rosen:1968mfz}, T. D. Lee~\cite{Friedberg:1976me}, and S. Coleman~\cite{Coleman:1985ki} and their collaborators.
Examples of NTSs exist in the Minimal Supersymmetric Standard Model (MSSM) where there are nonlinear interactions between squarks or sleptons which carry baryon or lepton numbers~\cite{Kusenko:1997zq,Kusenko:1997si}.
In Higgs-portal dark matter models, there may exist NTSs of the dark scalar fields inside which the electroweak symmetry is restored even after the electroweak phase transition~\cite{Ponton:2019hux}.
An NTS can also have gauge charges~\cite{Lee:1988ag,Gulamov:2015fya,Brihaye:2015veu,Heeck:2021zvk,Heeck:2021bce} or even topological charges~\cite{Bai:2021mzu} in the presence of a gauge group. Other studies include \cite{Kusenko:1997ad,Dvali:1997qv,Kusenko:1997vi,Berkooz:2005sf,Bishara:2017otb,Heeck:2020bau,Bai:2021xyf,Almumin:2021gax,Lennon:2021zzx,Pearce:2022ovj}; see \cite{Lee:1991ax,Nugaev:2019vru} for reviews.

Cosmologically, NTSs can be formed through a first- or second-order phase transition (FOPT or SOPT, respectively)~\cite{Frieman:1988ut,Griest:1989cb,Frieman:1989bx,Macpherson:1994wf,Hong:2020est} 
sometimes referred to as ``solitogenesis,''
where the matter in the unbroken phase is congregated by the bubble wall and condenses as small pockets.
In this circumstance, the two states of the same scalar field, the free scalar particles in the broken phase and the NTSs in the unbroken phase, will co-exist.
The interactions between free particles and NTSs could then possibly drive the cosmic properties of the NTSs away from ``the initial condition" right after the phase transition (PT).
In other words, the NTSs may absorb or release free scalar particles, or be formed through fusion of the free particles during the cosmic evolution, such that the typical charge and mass of the NTS system evolves with time. This is known as ``solitosynthesis.''
The properties of NTSs depend strongly on the interactions, and therefore the results of solitosynthesis can be very different between models.

In this work, we study the cosmic evolution of NTSs with and without solitosynthesis for different globally charged scalar models. 
Assuming an initial global charge asymmetry, we set up Boltzmann equations for the free-particle--NTS system and examine which species dominates the total charge or energy of the system.
This enables the estimation of two crucial temperatures of solitosynthesis: the NTS {\it charge-domination temperature} $T_D$ when the charge abundance of NTSs dominates that of free particles, and the {\it freeze-out temperature} $T_F$ when the NTS number density goes out of equilibrium. 
The former temperature being higher renders NTSs to be the dominant component of charge in the dark sector, which we refer to as ``efficient'' solitosynthesis.
For a given minimum stable charge $\Qmin$ and maximum attainable charge $\Qmax$, the minimum global charge asymmetry for efficient solitosynthesis follows a universal scaling relationship with very little model dependence. This degeneracy of parameter space is broken when the relic abundance of the dark sector is taken into account, as the mass spectrum of NTSs is model dependent. The upshot is that there exists a range of model parameters where NTSs can dominate the charge and/or energy density of the dark sector and be all of dark matter.

The cosmic evolution of macroscopic states in general has been discussed in several earlier works. Among them, Ref.~\cite{Griest:1989bq} coined the term and pioneered the study of ``solitosynthesis'' (see also \cite{Frieman:1989bx,Kusenko:1997si,Postma:2001ea}).
One finding of our work is that Ref.~\cite{Griest:1989bq} used an inappropriate estimate for the freeze-out temperature for solitosynthesis that leads to inaccurate estimates for the typical NTS charge, mass, and abundance. Their estimate relies on the freeze out of free particle annihilation, whereas our updated estimate relies on the freeze out of the NTS number density. 
Additionally, we find that the maximum NTS charge accessible during solitosynthesis is underestimated by Ref.~\cite{Griest:1989bq}, leading them to exclude far too pessimistically the possibility of NTS domination from solitosynthesis. Further, unlike prior works we distinguish between the charge domination and energy density domination of solitons.

Though we include only scalar NTSs in this paper, our discussions can be easily generalized to fermionic macroscopic states.
Examples of these include dark nuclei~\cite{Wise:2014jva,Wise:2014ola,Gresham:2017zqi,Gresham:2017cvl}, (dark) quark nuggets~\cite{Bai:2018vik,Bai:2018dxf,Liang:2016tqc}, and fermi-balls/fermion solitons \cite{Lee:1986tr,Macpherson:1994wf,Hong:2020est}. In Ref.~\cite{Gresham:2017cvl}, the coagulation of dark nuclei---heavy states of fermionic dark nucleons described by the liquid drop model whose mass and radius spectra in terms of the dark nucleon number are very similar to those of NTSs---is discussed and provides a rough analog to solitosynthesis.
Despite some model similarities, we find that the NTS evolution proceeds in a qualitatively different way, mostly relying on absorption of free particles as in \cite{Griest:1989bq} instead of mergers of larger bound states as in \cite{Gresham:2017cvl}. Still, like \cite{Gresham:2017cvl}, we find that states with large charges are formed.

On the other hand, if solitosynthesis is not efficient---which may occur when the free particles are out of equilibrium from NTSs, \eg, when the free-particle mass is much heavier than the PT temperature---then NTSs do not appreciably evolve after the PT. In this scenario, we examine how the PT parameters determine the typical charge, mass, and abundance of the NTSs, building on prior works. 

The organization of this paper is as follows. 
We first review several different scalar theories containing NTSs in Sec.~\ref{sec:Qball-model}. 
In Sec.~\ref{sec:solitosynthesis}, we study the solitosynthesis for two different NTS models and derive the parameter space where solitosynthesis is efficient and reproduces the observed dark matter relic abundance. Then, in Sec.~\ref{sec:Q_ball_from_PT}, we explore the model parameter space in the absence of solitosynthesis, where the cosmic properties of the dark sector are determined by the PT.
Finally, we discuss the related phenomenology and conclude in Sec.~\ref{sec:conclusion}. 
Several detailed calculations for the NTS properties and PTs are included in Appendices.

Throughout the paper we use the reduced Planck mass $\Mpl=1/\sqrt{8\pi G} = 2.44 \times 10^{18}~\GeV$ with $G$ the Newton constant.
Also, for simplicity we will follow the convention of~\cite{Coleman:1985ki} and use the terms ``Q-balls'' and ``nontopological solitons'' interchangeably.

\section{Q-ball models and properties}
\label{sec:Qball-model}

\subsection{Representative models}\label{sec:EWSDMB}

A general Q-ball state requires the theory to have a good global symmetry and a nonminimal potential~\cite{Coleman:1985ki}.
The mass per charge for a stable Q-ball state is smaller than a free particle carrying a unit charge in the vacuum of the potential. 
In this paper, we use a scalar boson constituent as a representative model and consider two renormalizable potentials for a theory of two gauge-singlet scalar fields. The first model, from Ref.~\cite{Frieman:1988ut}, is the one adopted in an earlier study of solitosynthesis~\cite{Griest:1989bq} with a real scalar $\sigma$ and complex scalar $S$ with scalar potential
\beqa
V(S, \sigma) = \frac{1}{8} \lambda\, (\sigma^2 - \sigma_0^2)^2 + \frac{1}{3} \lambda_2\, \sigma_0 (\sigma-\sigma_0)^3 +  \frac{m_S^2}{(\sigma_- - \sigma_0)^2} |S|^2 (\sigma - \sigma_0)^2 + \Lambda \, ,
\label{eq:VGK}
\eeqa
where $\sigma_-$ is the true zero-temperature minimum, 
$\Lambda$ is set so that $V=0$ at the zero-temperature minimum ($\sigma=\sigma_-$ and $S=0$), and for simplicity $\lambda_2=0.15 \lambda$ is used. The free-particle mass in the true vacuum is $m_S$. The quartic term $|S|^4$ was explicitly chosen to be zero in this model. We will call this ``Model A.'' 

This model admits a Q-ball solution for the field $S=e^{-i \omega t}\,\sigma_0\,s(r)/\sqrt{2}$ with $s(0)=s_0$ and $s(\infty)=0$, where $\sigma'(0)=0$ and $\sigma(\infty)=\sigma_-$. Its charge is $Q=i \int d^3 x \, \left(\Sbar\, \partial_t\, S - S\, \partial_t \,\Sbar \right) = 4 \pi \,\Omega \int_{0}^{\infty} d\rb \, \rb^2 s^2$, with $\Omega \equiv \omega/\sigma_0$ and $\rb \equiv \sigma_0\,r$. Q-balls in this model have mass $m_Q=4 \pi \sqrt{2} \,Q^{3/4} \Lambda^{1/4}/3$. By requiring their binding energy $B_Q=Q m_S - m_Q > 0$, it can be shown that the minimum stable charge is $\Qmin=1231 \Lambda m_S^{-4}$. Thus, the Q-ball mass and radius are,
\begin{align}
\label{eq:M_GK}
m_Q &= 5.15 \sigma_0 \lambda^{1/4} Q^{3/4} \, ,
\\
\label{eq:R_GK}
R_Q &= 0.8 \lambda^{-1/4} \sigma_0^{-1} Q^{1/4} \, .
\end{align}
The free particle mass in terms of these free parameters is $m_S=5.15 \sigma_0 \lambda^{1/4} \Qmin^{-1/4}$.

The second model, from Ref.~\cite{Ponton:2019hux}, has two complex scalars with scalar potential
\beqa
V(S, \phi) = \frac{1}{4} \lphi (|\phi|^2 - v^2)^2 +  \frac{1}{4} \lc |S|^2 |\phi|^2 + \ls |S|^4 + \ms^2 |S|^2 \, ,
\label{eq:V}
\eeqa
with $\lphi, \lc, \ls > 0$.
There are two global symmetries for this potential $U(1)_S$ and $U(1)_\phi$, where the $U(1)_S$ is responsible for the Q-ball charge. The other complex field $\phi$ is introduced to provide a nontrivial potential for the $S$ field. 
Similar results would be obtained if $\phi$ were a real scalar with $\mathbb{Z}_2$ parity ($\phi \rightarrow - \phi$) symmetry or a scalar multiplet under a larger symmetry group (including even $\phi$ being the Standard Model Higgs boson doublet as in \cite{Ponton:2019hux}). The zero-temperature free $S$ particle mass is $m_S^2 = \frac{1}{4} \lc v^2 + \ms^2$, and for simplicity we work in the limit $\ms=0$. We will call this ``Model B.'' The main difference between Model A and B is whether the self-quartic coupling for the $S$ field is zero or not.

This model also admits a Q-ball solution for the field $S=e^{-i \omega t}\,v\,s(r)/\sqrt{2}$ with $s(0)=s_0$ and $s(\infty)=0$, where $\phi = v\, f(r)$ satisfies $f'(0)=0$ and $f(\infty)=1$, provided $\lc^2 > 2 \lphi \ls$~\cite{Ponton:2019hux}. For sufficiently small $Q$, the $\ls$ term is negligible because the solutions have small $s_0$. Therefore, for small $Q$ the profile solutions can be approximated by $s(r) = s_0[1- \tanh^2(\omega' r)]$ and $f= 1 - \pi_0[1- \tanh^2(\omega' r)]$ for the case with $\pi_0 \ll 1$ ($\omega'$ is a parameter that is determined by minimizing the energy of the solution; see Appendix~\ref{appedix:smallQ} for details). The mass as a function of $Q$ is derived to be 
\beqa
\label{eq:M_small}
m_{Q, {\rm small} }\approx \left( \frac{1}{2}\,\lc^{1/2}\,+ \lphi\,\lc^{-1/2} \right) \, v\, Q - \frac{(\pi^2 - 9)}{2048\,\pi^2(\pi^2 - 6)^3}\,\lc^{5/2}\,v\,Q^3 ~.
\eeqa
The radius of the profile is $R_{Q, \rm small} \equiv \frac{1}{\omega'} \approx 64\pi(\pi^2 - 6)/(\lc^{3/2}\,v\,Q)$.

To be stable against decay to $Q$ free particles, $m_Q< Q m_S= Q\,\lc^{1/2} v/2$. The minimum (quantum-level) stable charge is
\beqa
\label{eq:Qmin}
Q_s \approx 32\pi\,\left( \frac{2(\pi^2-6)^3}{\pi^2-9} \right)^{1/2}\,\frac{\lphi^{1/2}}{\lc^{3/2}}\, .
\eeqa
In practice, $Q_s$ must be determined by numerically solving the classical equations of motion for $f(r)$ and $s(r)$. There is an additional charge $Q_c<Q_s$ which is the lowest possible Q-ball charge. Solutions do not exist for $Q<Q_c$, and in between $Q_c<Q<Q_s$ the Q-balls are metastable \cite{Levkov:2017paj}. The minimum (meta)stable quantized charge $\Qmin$ available during solitosynthesis should satisfy $\lceil Q_c \rceil \leq \Qmin \leq \lceil Q_s \rceil$, where metastability is defined by comparing to time scales relevant for solitosynthesis and $\lceil ... \rceil$ denotes the ceiling function. 
As we will see in the next section, a small $\Qmin$ would be preferred for efficient solitosynthesis, and hence the $Q_s$ of our concern will be of $\mc{O}(1)$, a value where its difference with $Q_c$ may be less than an integer and unimportant.

For quartic couplings of order unity, $\lambda_\phi = \mathcal{O}(1)$ and $\lambda_{\phi S} = \mathcal{O}(1)$, the minimum stable charge has $\Qmin = \mathcal{O}(10^3)$. On the other hand, for a flatter $\phi$ potential with $\lambda_\phi \ll 1$, a smaller $\Qmin$ is anticipated. For $\lambda_\phi$ saturating a minimum value from a one-loop n\"aive dimension analysis, $\lambda_\phi \sim \lc^2/(16\pi^2)$, one has $\Qmin = \mathcal{O}(1)$ from \eqref{eq:Qmin}.  

At large charge when $\ls>0$, the relationship between mass, radius, and charge are calculated to be
\begin{align}
    \label{eq:M_large}
    m_{Q, {\rm large}} & \approx  v\,Q \left[\left(\ls \lphi\right)^{1/4} + c_2 Q^{-1/3} \right]\, , 
    \\
        \label{eq:R_large}
    R_{Q, \rm large} & \approx \frac{3^{1/3} \, \ls^{1/12} }{(4\pi)^{1/3}\lphi^{1/4}\,v}\,Q^{1/3}\, .
\end{align}
We include the subleading  $Q^{2/3}$ surface energy term for the Q-ball mass with $c_2$ as a dimensionless number depending on couplings in the potential (see Appendix \ref{appedix:surfaceE} for derivation and the approximate formula of $c_2$). 
A nonzero repulsive self-interaction $\ls$ plays an important role in the parametrics at sufficiently large $Q$. 
If $\ls$ were exactly zero, then $m_Q \propto Q^{3/4}$ and $R_Q\propto Q^{1/4}$ at large $Q$ \cite{Ponton:2019hux}, the same scaling as in Model A. However, such a $\ls$ term must be present as it is generated by radiative corrections---\eg, in this theory, it generically satisfies $\ls \gtrsim \lc^2/(16 \pi^2)$ without fine tuning.

\begin{figure}[t!]
    \centering
    \includegraphics[width=0.46\textwidth]{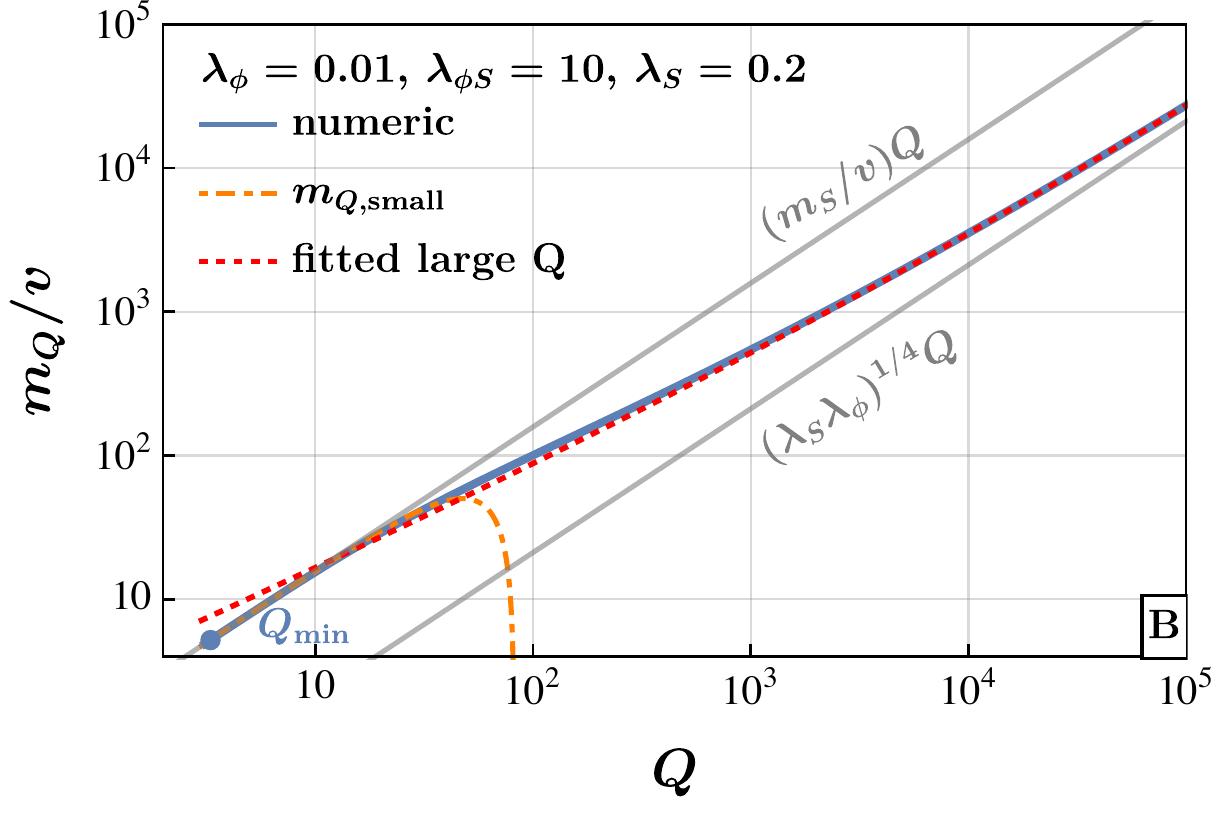} \hspace{4mm}
        \includegraphics[width=0.48\textwidth]{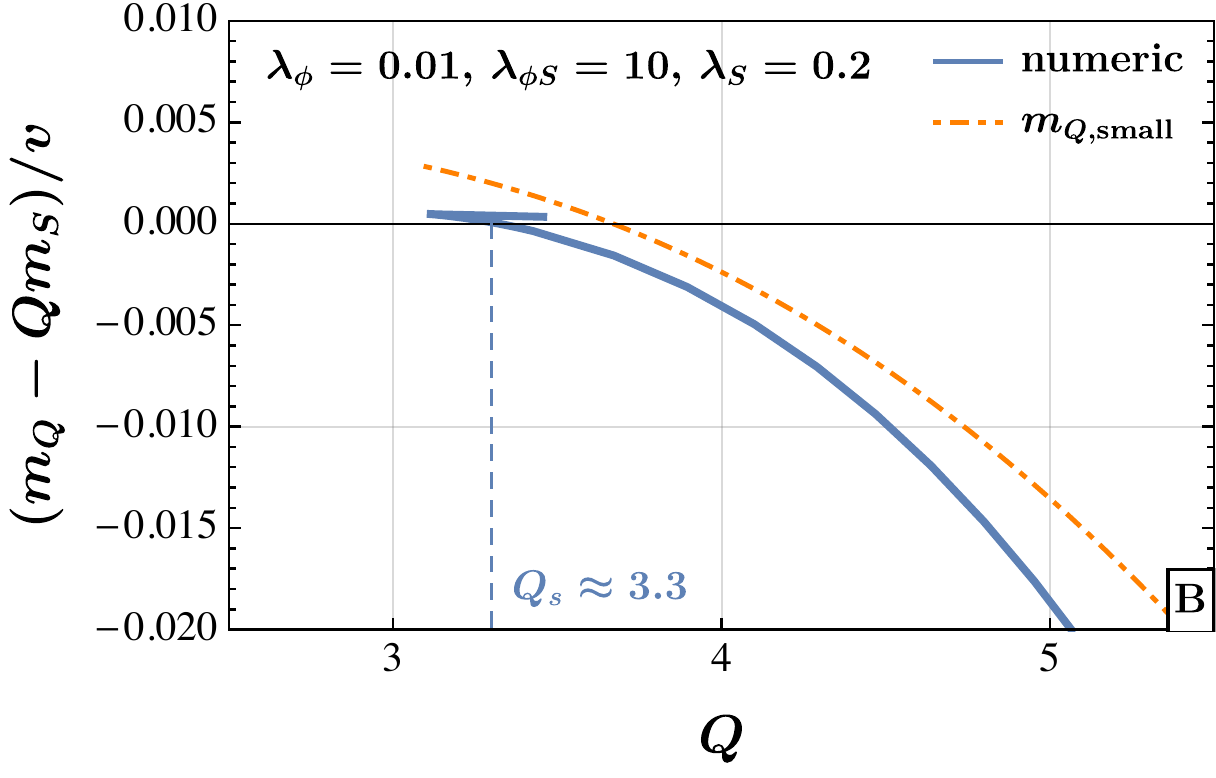}
    \caption{{\it Left panel:} The Q-ball mass as a function of charge $Q$ in Model B. The upper gray line shows the total mass of $Q$ free $S$ particles. The lower gray line shows the leading $Q$ dependence of the Q-ball mass in the large-$Q$ limit.
    The numerically obtained Q-ball mass spectrum is fitted with a linear term of $Q$ plus a $Q^{2/3}$ term as in \eqref{eq:M_large}, and the fit result is given as the red dotted line.
    The fitted coefficient for the $Q$ term matches well with $\left(\ls \lphi\right)^{1/4}$.
    {\it Right panel:} The zoom-in region for small $Q$, showing the mass difference of the Q-ball and $Q$ free $S$ particles as a function of $Q$. A negative value means that the Q-ball is stable against decaying into $Q$ free $S$ particles. 
     }
    \label{fig:MQ-Q}
\end{figure}

In Fig.~\ref{fig:MQ-Q}, we show the Q-ball mass as a function of $Q$ in Model B, where we fix the model parameters to be $\lambda_\phi = 0.01$, $\lambda_{\phi S} = 10$, and $\lambda_S = 0.2$. In the left panel, the numerically calculated results agree well with the analytic formulas in \eqref{eq:M_small} and \eqref{eq:M_large} for the small $Q$ and large $Q$ regions, respectively. In the right panel, we zoom in on the behaviors in the small $Q$ region and demonstrate that a Q-ball is quantum-level stable if $Q \ge Q_{\rm min} = 4$ (which is well approximated by (\ref{eq:M_small}) and (\ref{eq:Qmin})). A higher-energy Q-cloud solution is also displayed in the right panel, which is truncated in the figure but should extend to $Q=\infty$. We will not consider Q-clouds further in this work (see Refs.~\cite{ALFORD1988323,Nugaev:2015rna,Levkov:2017paj} for relevant studies), though it is possible that some Q-balls could initially form as Q-clouds and then later relax to lower-energy Q-balls or evaporate to free particles, affecting fusion and capture cross sections discussed in the following subsection.

There are other types of models which can generate a small enough $Q_{\rm min}$. 
One example is related to the Coleman-Weinberg potential~\cite{Coleman:1973jx} for the $\phi$ field. 
Instead of a tree-level potential for $\phi$, the self-interacting potential could be replaced by
\beqa
V(S, \phi) \supset \lambda_\phi \, |\phi|^4 \left[ 4 \log\left( \frac{|\phi|}{f}\right) - 1  \right] ~,
\eeqa
where $\lambda_\phi>0$ is loop-factor suppressed and the constant 4 in front of the logarithm is chosen to have the $\phi$ vacuum expectation value $\langle |\phi|\rangle  = f$. 
A second example is motivated by the flat directions (with vanishing tree-level $F$-term and $D$-term potential) in the MSSM~\cite{Dine:1995kz}. For instance, the  phenomenological potential in the $u^c d^c d^c$ direction is
\beqa
V(S, \phi) \supset  m_{\phi,0}^2 \left[ 1 - \tilde{c}_1\,\frac{\alpha_s}{8\pi}\,\frac{M_{\tilde{g}}^2 }{m_{\phi,0}^2} \,\log\left(\frac{M^2_{\tilde{g}} + \tilde{c}_2\,g_s^2\, |\phi|^2 }{M_X^2} \right) \right]|\phi|^2  + \frac{|\phi|^{2d}}{\Lambda^{2d -4 }} ~.
\eeqa
Here, $\tilde{c}_1$ and $\tilde{c}_2$ are order-one numbers related to $SU(3)_c$ group representations; $M_{\tilde{g}}$ is the gluino mass; $M_X$ is a high reference scale to define the soft masses $m_{\phi,0}$ and $M_{\tilde{g}}$. The last higher-dimensional operator with $d > 2$ makes the potential bounded from below. 

In the remainder of our paper, we use Model A in \eqref{eq:VGK} and Model B in \eqref{eq:V} as representative models to discuss early-universe formation of NTSs. Unless stated otherwise, in all figures we use the benchmark parameters: 
\begin{subequations} \label{eq:benchmark}
\begin{align}
    \label{eq:benchmark-A}
    \lambda=1 \, , \lambda_2=0.15 \lambda \, ,  \; \; \; \; & \text{(Model A)}~,
    \\
    \lambda_\phi=0.01\, , \lambda_{\phi S}=10\, , \lambda_S=0.2\, , \ms=0 \, ,  \; \; \; \; &  \text{(Model B)}~.
\end{align}
\end{subequations}
For Model A, $\Qmin$ remains a free parameter which determines $m_S$, while for Model B, these parameters fix $\Qmin=4$. It can be numerically confirmed that decreasing $\lc$ or $\ls$ or increasing $\lphi$ or $\ms$ will result in an increase to $\Qmin$. Also, we may at times use $\sigma_0$ and $v$ interchangeably.

\subsection{Q-ball interactions}
\label{sec:Qball_interactions}

Various processes could be important for the evolution of Q-balls with different charges. For processes with two initial states in the forward direction, one has
\begin{subequations}\label{eq:proc-main}
\beqa
\label{eq:proc-SS}
S + \Sbar &\leftrightarrow& \phi + \phi^\dagger ~, \\
\label{eq:proc-capture}
(Q) + S &\leftrightarrow& (Q+1) + X ~, \\
\label{eq:proc-Qsbar}
(Q) + \Sbar &\leftrightarrow& (Q-1) + X ~, \\
\label{eq:proc-Qmin}
(Q_{\rm min}) + \Sbar &\leftrightarrow&  \underbrace{S + S + \cdots +S}_{Q_{\rm min} - 1} + X ~. \\
\label{eq:proc-QQ}
(Q_1) + (Q_2) &\leftrightarrow& (Q_1 + Q_2) + X ~, \\
\label{eq:proc-QantiQ}
(Q_1) + (-Q_2) &\leftrightarrow& 
\left\{ \begin{array}{l l}
(Q_1 - Q_2) + X ~\, ~~ & \text{for} ~ Q_1 - Q_2 \ge Q_{\rm min} ~, \vspace{2mm} \\
\underbrace{S + S + \cdots +S}_{Q_1-Q_2} + X ~\, ~~ & \text{for} ~ \Qmin >  Q_1 - Q_2 \geq 0 ~.
\end{array} \right.
\eeqa
\end{subequations}
Here, $X$ represents the degrees of freedom in $\phi$, $\sigma$, or  the SM if their masses are smaller than the binding energy in the process. 
States with parentheses like $(Q)$ represent Q-balls with charge $Q$. All $Q$ are taken positive, while the equivalent processes for negative $Q$ can be inferred. For the first process in \eqref{eq:proc-SS}, the final state could be the radial or the Goldstone boson mode inside the complex field $\phi$ in Model B. In the limit of $m_S \gg \sqrt{\lambda_\phi} \,v$, the annihilation rate to both $\phi$ degrees of freedom is 
\beqa
\label{eq:SS-xsec}
\sigma v_{\rm rel} (S + \Sbar \rightarrow \phi + \phi^\dagger)  = \frac{1}{2^{11}\,\pi}\,\frac{\lc^2}{m_S^2} = \frac{1}{128\pi\,v^2} ~,
\eeqa
where $v_{\rm rel}$ is the relative velocity of the two initial-state particles. \footnote{An interesting case appears when $\phi$ is a real scalar field and more massive than $S$. If $S$ cannot kinematically annihilate to other particles, then an NTS can be made of both $S$ and $S^\dagger$ simultaneously \cite{Frieman:1989bx}.} 
In Model A the annihilation rate of the corresponding process is estimated as $\sigma v_{\rm rel}=0.014(1-\Qmin^{1/2}/12.5)^{1/2}/(\Qmin^{1/2}\sigma^2_0)$~\cite{Griest:1989bq}.

For a large $Q$, the elastic scattering cross section of $(Q) + S \rightarrow (Q) + S$ can be approximated by a geometric cross section, $\pi R_{Q, \rm large}^2$. For the capture process in \eqref{eq:proc-capture}, the cross section formula depends on how the binding energy is released into $X$ and is hence model dependent. The detailed calculation is beyond the scope of the current paper (see Ref.~\cite{Bai:2019ogh} for one example of a radiative capture cross section calculation; the current model has self-quartic interactions of the free $S$ particle with the $S$-constituents inside the Q-ball, which will change the capture cross section relative to \cite{Bai:2019ogh}).  
To simplify our discussion, we choose $\sigma v_{\rm rel}$ for the capture processes in (\ref{eq:proc-capture}) and (\ref{eq:proc-Qsbar}) to be the geometric cross section $\pi R_Q^2$ using (\ref{eq:R_GK}) for Model A and and $R_{Q, \rm large}$ in (\ref{eq:R_large}) for Model B.
We  will also apply these cross section formulas to the region close to $Q_{\rm min}$, although the cross section is unlikely to be geometric because only a handful of bound states mediate the scattering. 

For the process in \eqref{eq:proc-Qmin} and depending on the value of $Q_{\rm min}$ and the abundance of free particles, a detailed balance may be difficult to reach. The forward Q-ball destruction process could happen much more easily than the backward fusion process of free $S$ particles forming a Q-ball state. For the small $Q_{\rm min}$ region of interest in Sec.~\ref{sec:solitosynthesis}, detailed balance is easier to maintain. We will see in the later analysis that this process plays an important role in determining when the Q-ball number density goes out of equilibrium. The case where the fusion process is absent is considered in \cite{Griest:1989bq,Frieman:1989bx} with largely negative results for solitosynthesis.

The last two processes in \eqref{eq:proc-QQ} and \eqref{eq:proc-QantiQ} could be important to change the charge distribution of Q-balls, but not generally important to change the total $S$-number inside Q-ball or free particle states. They will not change the equilibrium distributions of the Q-balls (to be discussed in Sec.~\ref{subsec:TD}). Further, because Q-ball charges will tend towards $\Qmax$ (defined in Sec.~\ref{sec:maximum-charge}) during the later stages of solitosynthesis, these processes will not have much effect on the final charge distribution. There may be a small effect on the freeze-out temperature. Additionally, in \cite{Bai:2021mzu}, it was estimated (by comparing the interaction rate $\Gamma = n_Q \sigma v_\text{rel} \sim n_Q \pi R^2 \sqrt{T/m_Q}$ with the Hubble parameter $H$) that $(Q)+(Q)$ interactions are only important when $Q \lesssim 6 \times 10^4 (v/10^3~\GeV)^{-12/5}$ for the case of Model B near temperatures $T\sim v$ assuming those Q-balls make up all of dark matter. Notice that the rate is suppressed by both the Q-ball number density $n_Q$ and nonrelativistic velocity when $Q$ is large. A similar estimate for Model A gives $Q \lesssim 2\times 10^3 (\sigma_0/10^3~\GeV)^{-16/5}$. Because (i) this is well below $\Qmax$, and (ii) the Q-balls in this range typically have much smaller comoving density than the dark matter density throughout the course of solitosynthesis, we can safely neglect these interactions. When there is an asymmetry present, as will be required for solitosynthesis but not necessarily PT formation, \eqref{eq:proc-QantiQ} can also generally be neglected in the formation process because the abundance of anti-Q-balls is suppressed.

Relatedly, Ref.~\cite{Gresham:2017cvl} includes processes analogous to $(Q_1)+(Q_2)\leftrightarrow (Q_3)+(Q_4)$ in the evolution, where $Q_1+Q_2=Q_3+Q_4$.
However, it is also argued in~\cite{Gresham:2017cvl} that under the on-shell scheme, in the process $(Q_1)+(Q_2)\to (Q_1+Q_2)^\ast\to (Q_3)+(Q_4)$ the intermediate excited state $(Q_1+Q_2)^\ast$ prefers to decay to $(Q_1+Q_2) + X$. We expect similar behavior in our models. Therefore the net effect of this process will be similar to the combination of our \eqref{eq:proc-capture} and \eqref{eq:proc-QQ}, and other charge combinations in the final state can generally be neglected.

\section{Q-balls from solitosynthesis}
\label{sec:solitosynthesis}

Cosmologically, Q-balls may be conveniently formed from PTs, while their final abundances may not necessarily be determined merely by the PT.
If there exists an efficient solitosynthesis, wherein Q-balls can form from the merger of $\Qmin \times S$ particles and grow or shrink from Q-balls absorbing or emitting $S/\Sbar$ particles, the initial abundance of Q-balls after the PT would be irrelevant. 
The interactions within the dark sector would generate an equilibrium system up to a certain large charge and finally decouple from the thermal bath due to the cosmic expansion.
Thus, to know the final properties and the relic abundance of the dark sector, one needs to track its evolution by examining the solitosynthesis process, which we examine in this section.

The $U(1)_S$ charge asymmetry $\eta$ plays an important role in the course of solitosynthesis. 
If $\eta=0$, the Q-ball abundance is quickly depleted in equilibrium and the free $S$ particles dominate the global charge.
Thus, for the purposes of this section, we assume $\eta>0$ and check how small $\eta$ can be to have Q-balls dominate the global charge.

\subsection{The maximum charge in the Q-ball system}
\label{sec:maximum-charge}
Assuming that the solitosynthesis processes are not immediately frozen out after the PT, Q-balls of all possible charges could be successively produced and reach a thermal distribution through direct fusion or absorption/emission of free quanta of the global charge.
As the universe cools down, the global charges will gradually congregate into the Q-balls due to the gain of binding energy.
Similar to the analysis in~\cite{Griest:1989bq}, we can determine the temperature at which the majority of the global charges are in Q-balls, but to do this we need to understand the maximal Q-ball global charge accessible during solitosynthesis.

For Models A and B presented in Sec.~\ref{sec:Qball-model}, there is no intrinsic upper bound on the Q-ball charge.
Nevertheless, having an upper bound $\Qmax$ on the charge of the system is not only necessary for numerical computations, but also realistic in terms of statistical equilibrium, since it takes time for the Q-balls not produced from the PT to be formed and thermalized.
In statistical equilibrium, $\Qmax$ can be estimated by comparing the time required for a Q-ball to absorb that many $S$ particles and the Hubble time. For statistical equilibrium to hold, we expect that a Q-ball should be able to charge from $\Qmin$ to $\Qmax$ within a few Hubble times. The charge time is
\begin{equation}
\label{eq:charge-shuffle-time}
    \tau_{\Qmin \to \Qmax} = \sum_{Q=\Qmin}^{\Qmax} \frac{1}{n_S \, (\sigma v_{\rm rel})_Q} \, ,
\end{equation}
where $n_S$ is the number density of $S$ particles. For Model A where $R_Q \simeq 0.8 (Q/\lambda)^{1/4}/\sigma_0$, the summation can be approximated [using the geometric cross section $(\sigma v_{\rm rel})_Q \approx \pi R_Q^2$] as $\tau_{\Qmin \to \Qmax} \simeq \lambda^{1/2} \sigma_0^2  \Qmax^{1/2} / n_S$. Requiring $\tau_{\Qmin \to \Qmax} \lesssim H^{-1}$ with $H=\sqrt{\pi^2 g_*/90}\, T^2/\Mpl$,  
\begin{equation}\label{eq:Qmax_bound}
    \Qmax \lesssim \left( \frac{3\,n_{S}\, \Mpl}{g^{1/2}_\ast\,\lambda^{1/2}\, \sigma_0^2\, T^2} \right)^2 \,,
\end{equation}
where $g_\ast$ is the relativistic degrees of freedom in the thermal plasma.
For Model B, the corresponding upper limit is
\begin{align}\label{eq:Qmax_bound_2}
    \Qmax\lesssim \left(\dfrac{3^{2/3} \ls^{1/6} }{(4\pi)^{2/3}\lphi^{1/2}}\dfrac{\pi\uu n_S \Mpl}{g^{1/2}_\ast v^2 T^2}\right)^3\,,
\end{align}
using the radius in \eqref{eq:R_large}. These limits are less constraining for a smaller energy scale $\sigma_0$ or $v$.

A similar condition for an initial population of Q-balls following the PT to discharge in order to reach equilibrium could be considered by replacing $n_S$ with $n_{\Sbar}$, the number density of $S^\dagger$. The maximum charge for an NTS that is capable of being discharged down to $\Qmin$ should not differ much from the bound on $\Qmax$ estimated above. This is because $S S^\dagger$ pair creation in the reverse process of (\ref{eq:proc-SS}) is still active at temperatures $T \sim v$ or $\sigma_0$ right after the PT, so $n_S \sim n_{S^\dagger}$.

Note that the estimation of $Q_{\rm max}$ here is different from Ref.~\cite{Griest:1989bq}, where the Q-ball number density $n_{\Qmin}$ is used in place of $n_S$ to estimate the charge shuffling time in the right-hand side of (\ref{eq:charge-shuffle-time}). We have also numerically checked the time for a perturbed system of number densities to return to equilibrium and found agreement with Eq.~\eqref{eq:charge-shuffle-time}. 

\subsection{Q-ball domination in equilibrium}
\label{subsec:TD}

The number density of the free global charge quanta $S$, $\Sbar$ and NTSs of charge $Q$ and $-Q$ in kinetic equilibrium are given by $n_{i=S,\Sbar,Q,-Q}^{\text{eq}}$, with
\begin{align}\label{eq:number_density}
    n_i^{\text{eq}} = \frac{1}{2\pi^2}T\,m^2_i\exp(\mu_i/T)K_2(m_i/T)\,,
\end{align}
where $K_2(x)$ is a modified Bessel function. In chemical equilibrium, the chemical potentials satisfy [based on (\ref{eq:proc-main})]
\begin{subequations}
\label{eq:mu}
\begin{align}
    \mu_{\Sbar} &= -\mu_S\,,\\
    \mu_Q &= Q\,\mu_S\,,\\
    \mu_{-Q} &= -\mu_{Q}\,.
\end{align}
\end{subequations}
If $\eta=0$, the chemical potentials are identically zero, and the Q-ball abundances are Boltzmann suppressed such that solitosynthesis is inefficient. In the following, we take $\eta>0$.
Letting $Z_i\equiv n_i/(\eta \, n_\gamma)$, the conservation of global charge indicates
\begin{align}\label{eq:Q_conservation}
    Z_S-Z_{\Sbar}+\sum^{Q_{\rm max}}_{Q=Q_{\rm min}}Q\,(Z_{Q}-Z_{-Q}) = 1\, .
\end{align}
For a given mass spectrum $m_i$ and asymmetry $\eta$, this equation uniquely determines the chemical potential $\mu \equiv \mu_S$ as a function of temperature.

For models of interest to us, the global charge will be mainly aggregated in Q-balls of charge $\Qmax$ at low temperature as long as the Q-ball system stays in equilibrium. 
To see this, we parametrize the mass spectrum of the Q-ball system to be $m_Q=m_1 Q^p$. We expect $p\leq 1$, such that $m_Q/Q$ will be smaller than the free charge quanta mass for sufficiently large $Q$.
Defining a ratio $r=(Q+1)n_{Q+1}/(Q\,n_Q)$ with \eqref{eq:number_density} and using $K_2(x)\approx \sqrt{\pi/(2x)}e^{-x}$ for $x \gg 1$, 
\begin{align}
\label{eq:rQ}
    r&=\left(\dfrac{Q+1}{Q}\right)^{\frac{3p}{2}+1}\exp\left(\dfrac{m_Q-m_{Q+1}+\mu}{T}\right)\,,\\
\label{eq:drdQ}
    \frac{dr}{dQ}&=\frac{2 p\,m_1\left(Q^p(Q+1)-(Q+1)^p Q\right)-(2+3p)T}{2\,Q^2\, T}\left(\frac{Q+1}{Q}\right)^{3p/2}\exp\left(\dfrac{m_Q-m_{Q+1}+\mu}{T}\right)\,.
\end{align}
The sign of $dr/dQ$ is determined by the numerator of the first term $2p\,m_1(Q^p(Q+1)-(Q+1)^p Q)-(2+3p)T$. For $p=1$, $dr/dQ$ will be negative, therefore the charge of the equilibrium system will concentrate in NTSs with a specific charge $Q<\Qmax$ for sufficiently large $\Qmax$. For $p<1$, on the other hand, $r \xrightarrow[]{Q\to\infty} e^{\mu/T}>1$.
Therefore, as long as $\Qmax$ is large enough, the charge of the system will finally concentrate in the largest Q-balls $(\Qmax)$ following the equilibrium evolution.
Of course, we usually do not expect the Q-ball mass spectrum to have a simple power-law behavior, as we have seen in Sec.~\ref{sec:Qball-model}. However, as long as the growth of $m_Q$ versus $Q$ is ``slower'' than the linear power, one effectively has the $p<1$ scenario with $(\Qmax)$ finally dominating the global charge.

\begin{figure}[t!]
\centering
    \includegraphics[width=0.46\textwidth]{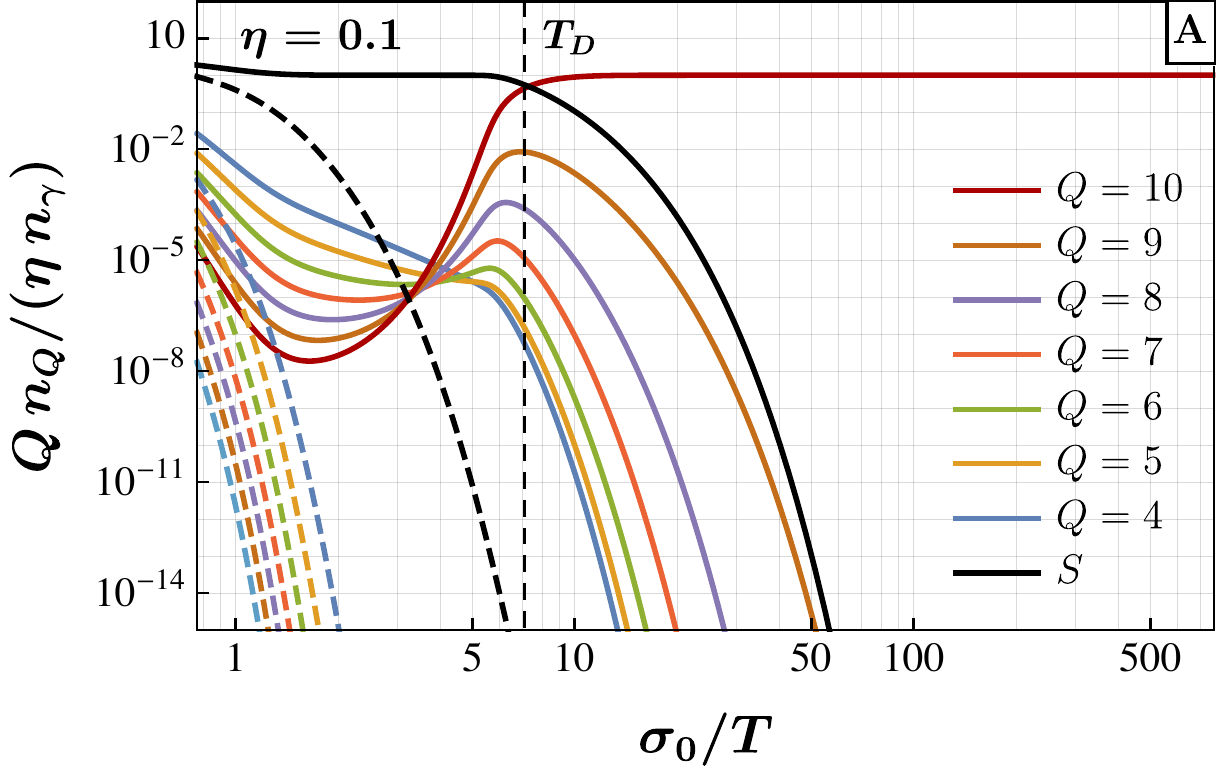} \hspace{4mm}
    \includegraphics[width=0.48\textwidth]{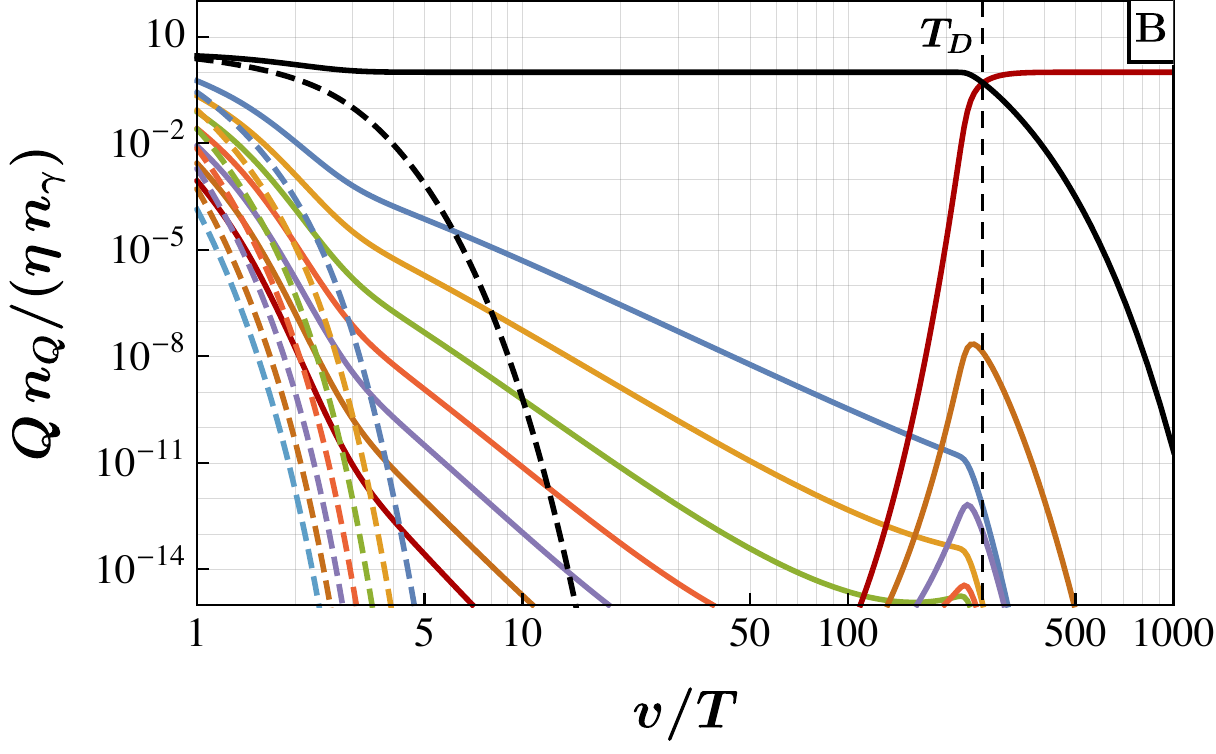}
    \caption{The evolution of the equilibrium distribution of global charge for Model A with mass spectrum in (\ref{eq:M_GK}) (left panel) and Model B with mass spectrum obtained numerically (right panel) using model parameters from (\ref{eq:benchmark}).
    For both panels, $\eta=0.1$, $\Qmin=4$, and $\Qmax=10$. The solid curves represent the evolution of $S$ and positively-charged solitons $(Q)$, while the dashed curves are for $\Sbar$ and negatively-charged solitons $(-Q)$. 
    The vertical dashed lines indicate $T_D$ as analytically estimated in \eqref{eq:TD}.
    }
    \label{fig:equilibrium}
\end{figure}

In Fig.~\ref{fig:equilibrium} we show the equilibrium evolution of the global charge for Models A and B. In both models, the global charge stored in the largest Q-ball dominates over those in other species at late time.
The overall shape of the curves in Fig.~\ref{fig:equilibrium} can be understood in the following way. At the earliest times, all interactions in (\ref{eq:proc-main}) are active and the chemical potential is near zero, favoring lighter states, \ie, free particles and smaller-charged Q-balls. Then, as temperature drops the reverse process in (\ref{eq:proc-SS}) becomes inefficient and the abundance of $S^\dagger$ decreases well below the abundance of $S$, so further annihilations via (\ref{eq:proc-SS}) do not appreciably change the $S$ abundance. The chemical potential increases, and anti-Q-ball abundances also become suppressed. Eventually it becomes kinematically inefficient to knock charges out of Q-balls in the reverse process of (\ref{eq:proc-capture}), so the only remaining processes are Q-ball fusion in the reverse of (\ref{eq:proc-Qmin}) and Q-ball captures in the forward process of (\ref{eq:proc-capture}). Because maximally-charged Q-balls are the lowest energy state per charge, they become the most abundant, leading to the bounce in their abundance at late times (as demonstrated in (\ref{eq:drdQ}), see also \cite{Puetter:2022ucx} for a discussion of this bounce in a simplified three-particle system).
The Q-ball domination in Model B is much later in the presented examples because at small $Q$ the mass spectrum in Model B is still dominated by the linear term in $Q$ and thus has a smaller binding energy, as shown by Eq.~\eqref{eq:M_small} and Fig.~\ref{fig:MQ-Q}.

To understand the equilibrium evolution of the system better, we estimate the temperature $T_D$ where the Q-ball charge domination happens as in \cite{Griest:1989bq}.
It is clear from \eqref{eq:mu} that $n_{\Sbar}$ and $n_{-Q}$ are suppressed by the chemical potential term compared with $n_S$ and $n_{Q}$.
Therefore at $T=T_D$ we expect $Z_S \approx \Qmax Z_{\Qmax} \approx 1/2$.
By approximating $K_2(x)\approx \sqrt{\pi/(2x)}e^{-x}$ for $x \gg 1$ in~\eqref{eq:number_density}, 
\begin{align}
    &\left(\frac{m_S\,T_D}{2\pi}\right)^{3/2} \exp\left(\frac{\mu-m_S}{T_D} \right) = \Qmax \left(\frac{m_{\Qmax}\,T_D}{2\pi}\right)^{3/2} \exp\left(\frac{\Qmax\mu-m_{\Qmax}}{T_D}\right)=\frac{1}{2}\eta c_\gamma T_D^3\,,
\end{align}
where $\mu\equiv\mu_S$, $B_{\Qmax}=\Qmax\,m_S-m_{Q}$, and $c_\gamma = 2\zeta(3)/\pi^2$.
Then,
\begin{align}\label{eq:TD}
    T_D=\frac{B_{\Qmax}}{\log\left\{\frac{1}{\Qmax}\left[\frac{2}{\eta c_\gamma}\left(\frac{m_S}{2\pi T_D}\right)^{\frac{3}{2}}\right]^{\Qmax-1}\left(\frac{m_S}{m_{\Qmax}}\right)^{\frac{3}{2}}\right\}}\,.
\end{align}
This analytic estimate is shown to match very well to the crossing of the $n_S^\text{eq}$ and $\Qmax n_{\Qmax}^\text{eq}$ curves in Fig.~\ref{fig:equilibrium}. As the system usually evolves to a very large $\Qmax$, one can obtain the asymptotic expression of $T_D$ in the limit of $\Qmax\to\infty$ for different models.
For the Q-balls in Model A, we parametrize the spectra $m_S=m_0\Qmin^{-1/4}$ and $m_{\Qmax}=m_0 \Qmax^{3/4}$ with $m_0 = 5.15 \sigma_0 \lambda^{1/4}$, and Eq.~\eqref{eq:TD} can be further rewritten as
\begin{align}
    T_D\xrightarrow[]{\Qmax\to\infty}\frac{m_0\,\Qmin^{-1/4}}{\log\left\{\frac{2}{\eta c_\gamma}\left(\frac{m_0}{2\pi T_D}\right)^{\frac{3}{2}}\,Q^{-3/8}_{\rm min}\right\}}\,, \qquad \mbox{(Model A)} ~.
\end{align}
For Model B, on the other hand, at large $\Qmax$ we expect the Q-ball mass to be $m_{\Qmax}=\Qmax\Omega_c v$ at the leading order, and the free global charge quanta mass to be $m_S=\sqrt{\lambda_{\phi S}}\,v/2$. Therefore, 
\begin{align}
\label{eq:TD-B}
    T_D\xrightarrow[]{\Qmax\to\infty} \frac{v\,(\sqrt{\lambda_{\phi S}}/2-\Omega_c)}{\log\left\{\frac{2}{\eta c_\gamma}\left(\frac{\sqrt{\lambda_{\phi S}}v}{4\pi}\right)^{\frac{3}{2}}T^{-\frac{3}{2}}_D\right\}}\,, \qquad \mbox{(Model B)} ~,
  \end{align}
where $\Omega_c \equiv (\ls \lphi)^{1/4}$. 

From the Q-ball domination we can also understand how the chemical potential $\mu$ evolves at  late time.
At $T<T_D$, we expect almost all the global charge to be concentrated in $(\Qmax)$, \ie, $n_{\Qmax}\simeq \eta\, n_\gamma / \Qmax$ using (\ref{eq:Q_conservation}), which gives an approximate analytic expression  of $\mu$:
\begin{equation}\label{eq:mu_analytic}
    \mu \simeq \frac{1}{Q_\text{max}} \left(m_{\Qmax} + T \log \left[\frac{\eta\,c_\gamma}{\Qmax} \left(\frac{2 \pi T}{m_{\Qmax}}\right)^{3/2} \right] \right) \, , \; \; \; \; T < T_D \, .
\end{equation}

A final remark before moving on to the out-of-equilibrium evolution of the system.
The temperature $T_D$ defined here is the temperature when the Q-balls dominate over the free particles in {\it charge}, but not necessarily {\it energy density}.
Setting the two energy densities equal, by approximating $n_{\Qmax} \simeq \eta\,n_\gamma / \Qmax$ and using \eqref{eq:mu_analytic}, the energy density domination temperature is  
\begin{align}\label{eq:Trho}
    \Trho=\dfrac{m_{\Qmax}-\Qmax m_S}{(\Qmax-1)\log\left[\dfrac{\eta\uu c_\gamma}{\Qmax}\left(\dfrac{2\pi\uu \Trho}{m_S}\right)^{3/2}\right]+\left(\Qmax+\dfrac{3}{2}\right)\log\left(\dfrac{m_{\Qmax}}{m_S}\right)}\,,
\end{align}
which in the large-$\Qmax$ limit becomes
\begin{align}\label{eq:Trho2}
    \Trho\xrightarrow[]{\Qmax\to\infty}\dfrac{m_S-m_{\Qmax}/\Qmax}{\log\left[\dfrac{\Qmax m_S}{m_{\Qmax}}\dfrac{1}{\eta\uu c_\gamma}\left(\dfrac{m_S}{2\pi\uu \Trho}\right)^{3/2}\right]}\,.
\end{align}
For Model A, $\Qmax/m_{\Qmax}\propto \Qmax^{1/4}$, and therefore Q-balls dominate the energy density at a rather late time for large $\Qmax$ compared to the charge-dominance temperature $T_D$.
For Model B, on the other hand, the situation is different. Because $m_{\Qmax}\sim \Qmax\, \Omega_c \,v$ for large $\Qmax$, $\Trho$ does not depend on $\Qmax$ and will be closer to $T_D$.
However, whether the Q-ball energy density domination can happen depends on when the system goes out of equilibrium, which we discuss in the next subsection.

\subsection{The freeze out and relic abundance of Q-balls}

The evolution of the free-particle--NTS system will have to freeze out sometime after the PT due to the cosmic expansion. In Ref.~\cite{Griest:1989bq}, the freeze-out temperature was estimated by considering the freeze out of process \eqref{eq:proc-SS}. However, we have numerically checked that this estimation does not capture the Q-ball dynamics properly because it is not directly related to Q-ball evolution.~\footnote{Ref.~\cite{Postma:2001ea} considered the freeze out of an individual species $n_Q$ rather than the sum $n_\text{NTS}$, which is not precise. One needs to account for $(Q-1)+S$ and $(Q)+S$ processes simultaneously, which adds to and removes from the abundance of $(Q)$ at similar rates. Ref.~\cite{Frieman:1989bx} largely disregards the fusion process in the reverse of (\ref{eq:proc-Qmin}).} Instead, one should examine the total Q-ball number density:
\begin{equation}
    n_\text{NTS} \equiv \sum_{Q=\Qmin}^{\Qmax} n_{Q} \, .
\end{equation}
We start with the Boltzmann equations of (\ref{eq:proc-capture}--\ref{eq:proc-Qmin}) for the individual Q-ball number densities,
\begin{equation}
\label{eq:boltz-single}
    \begin{aligned} 
    \dot{n}_{Q} + 3 H n_{Q} 
    = & 
    - \delta_{Q,Q_\text{min}} (\sigma v_{\rm rel})_{Q_\text{min}} \left(n_{\Qmin} n_{\Sbar} - n_{\Qmin}^\text{eq} n_{\Sbar}^\text{eq} \left( \frac{n_S}{n_S^\text{eq}} \right)^{\Qmin - 1} \right)
    \\ & 
    - (1-\delta_{Q,Q_\text{max}}) (\sigma v_{\rm rel})_{Q} \left(n_{Q} n_S - n_{Q}^\text{eq} n_S^\text{eq} \left( \frac{n_{Q+1}}{n_{Q+1}^\text{eq}} \right) \right)
    \\ & 
    + (1-\delta_{Q,Q_\text{min}}) (\sigma v_{\rm rel})_{Q-1} \left(n_{Q-1} n_S - n_{Q-1}^\text{eq} n_S^\text{eq} \left( \frac{n_{Q}}{n_{Q}^\text{eq}} \right) \right)
        \\ & 
    - (1-\delta_{Q,Q_\text{min}}) (\sigma v_{\rm rel})_{Q} \left(n_{Q} n_{\Sbar} - n_{Q}^\text{eq} n_{\Sbar}^\text{eq} \left( \frac{n_{Q-1}}{n_{Q-1}^\text{eq}} \right) \right)
    \\ & 
    + (1-\delta_{Q,Q_\text{max}}) (\sigma v_{\rm rel})_{Q+1} \left(n_{Q+1} n_{\Sbar} - n_{Q+1}^\text{eq} n_{\Sbar}^\text{eq} \left( \frac{n_{Q}}{n_{Q}^\text{eq}} \right) \right)
    \, ,
    \end{aligned}
\end{equation}
where $\delta_{i,j}$ are the Kronecker delta functions and $Q>0$ is assumed ($Q<0$ equations are obtained by swapping $S$ and $S^\dagger$). 
The cross sections are given by the Q-ball geometric cross section: $(\sigma v_\text{rel})_{Q} \approx \pi R_{Q}^2$ (although see discussion of this assumption in Sec.~\ref{sec:Qball_interactions}).~\footnote{Here, we have taken $v_\text{rel} \sim 1$ because the $S$ particles are semi-relativistic during solitosynthesis. We have verified that taking into account the additional velocity dependence does not change the result appreciably.}
All of the terms on the right-hand side of Eq.~(\ref{eq:boltz-single}) except the $(\sigma v_\text{rel})_{\Qmin}$ term are related to the internal evolution of the NTS system, but do not change $n_\text{NTS}$.
So, by summing the Boltzmann equations for each value $Q$ together, a simpler Boltzmann equation for the total Q-ball abundance is obtained via the cancellation of terms on the right-hand side:
\begin{equation}\label{eq:boltz-NTS}
    \dot{n}_\text{NTS} + 3\,H\,n_\text{NTS} 
    = 
    - (\sigma v_{\rm rel})_{Q_\text{min}} \left(n_{\Qmin} n_{\Sbar} - n_{\Qmin}^\text{eq} n_{\Sbar}^\text{eq} \left( \frac{n_S}{n_S^\text{eq}} \right)^{Q_\text{min} - 1} \right) \, .
\end{equation}
Thus, the freeze-out temperature of the NTS number density can be estimated from 
\begin{equation}\label{eq:TF}
    H\,n_\text{NTS} \sim (\sigma v_{\rm rel})_{Q_\text{min}}\, n_{\Qmin}\,n_{\Sbar}\,\left| \right._{T=T_F}  \, .
\end{equation}
Estimated in this way, $T_F$ provides a good proxy for comparison to $T_D$. 
This is because for $T<T_D$, the freeze out of $n_\text{NTS}$ is equivalent to the freeze out of the dominant charge component of the system $(\Qmax)$ because $n_\text{NTS}^\text{eq} \simeq n_{\Qmax}^{\rm eq}$ has already stopped significantly evolving (see Fig.~\ref{fig:equilibrium}). This approximate equality enables us to find a simplified expression for $T_F$ from~\eqref{eq:TF} by substituting the equilibrium number densities,
\begin{align}
\label{eq:TF2}
    T_F=\dfrac{(\Qmin-1-\Qmax)\mu-(m_S+m_{\Qmin}-m_{\Qmax})}{\log\left[\dfrac{\pi\,g^{1/2}_\ast \,T_F^{1/2}[2\pi\,m_{\Qmax} /(m_S\,m_{\Qmin})]^{3/2})}{\sqrt{90}\Mpl\,(\sigma v_{\rm rel})_{\Qmin}  } \right]}\, , \; \; \; \;\mbox{if}~T_F < T_D\,.
\end{align}

\begin{figure}[t!]
\centering
    \includegraphics[width=0.48\textwidth]{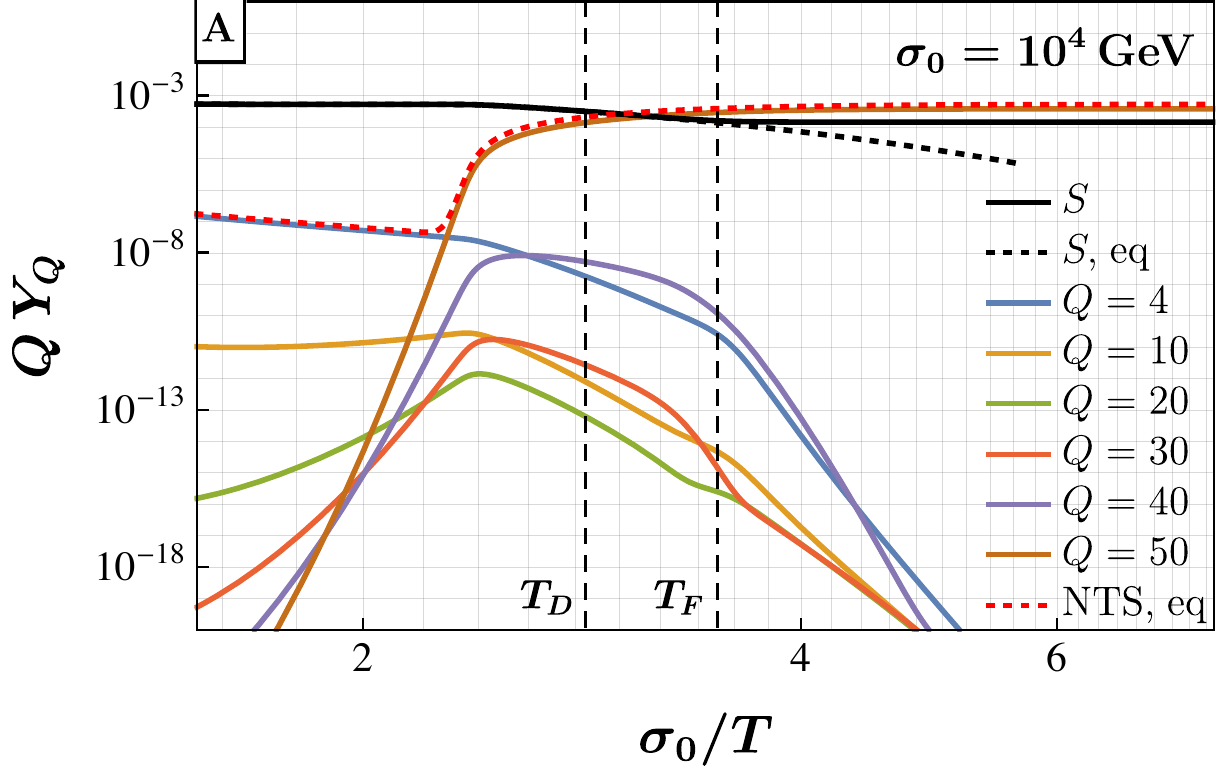} \hspace{4mm}
    \includegraphics[width=0.48\textwidth]{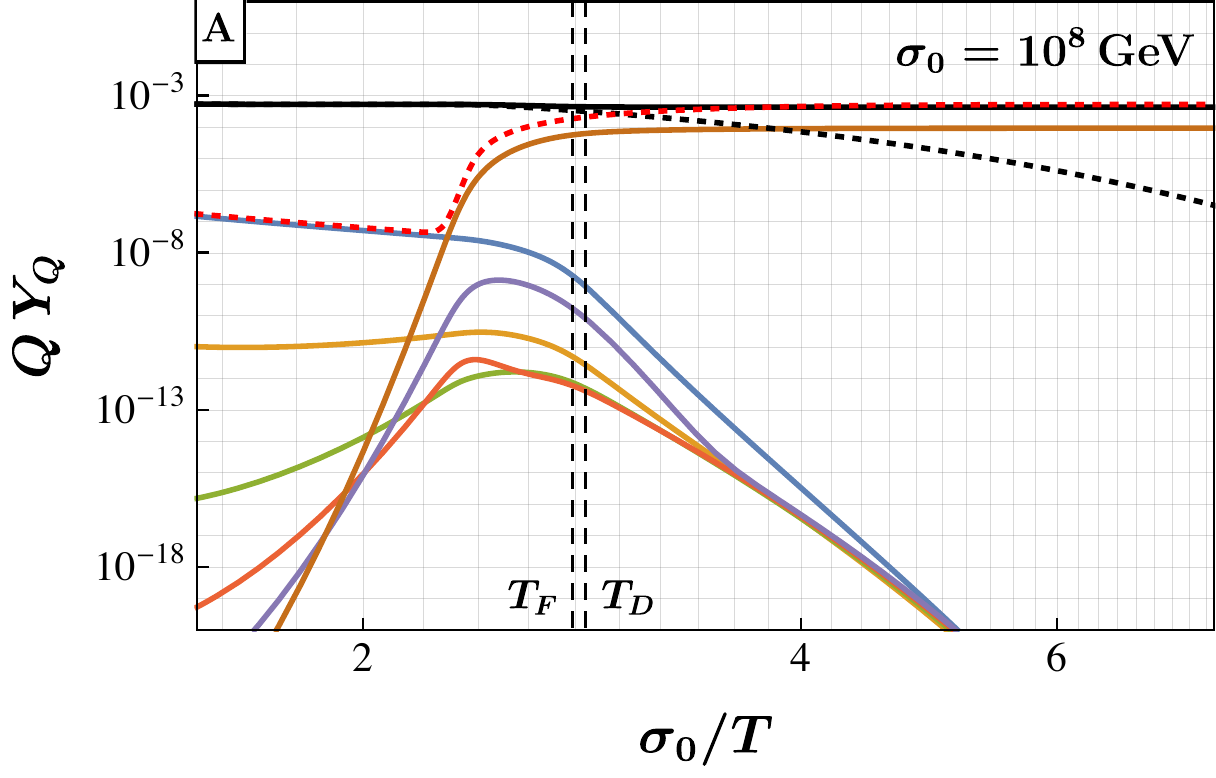}
    \caption{Evolution of charge yields $Y_Q \equiv n_Q/s$ for a few species in Model A after solving the full set of Boltzmann equations.    The left panel with $\sigma_0=10^4$\,GeV has Q-ball charge dominance, while the right panel with $\sigma_0=10^8$\,GeV has $S$ dominance.   Both plots use $\eta=0.1$, $\Qmin=4$, and $\Qmax=50$.
    The vertical dashed black lines represent the freeze-out temperature $T_F$ estimated from \eqref{eq:TF2} and the analytic expression for $T_D$ from \eqref{eq:TD} if chemical equilibrium were maintained. 
    The dotted black and red curves show the evolution of the equilibrium charge yields of the $S$ particle and NTS, respectively.
    }
    \label{fig:Boltzmann}
\end{figure}

In Fig.~\ref{fig:Boltzmann}, we show the numeric solutions to the full set of Boltzmann equations for two benchmark points of Model A,~\footnote{For Model B, it is not possible to probe large enough $\Qmax$ numerically to get a useful result where (\ref{eq:M_large}) determines the mass.} as well as the freeze-out temperature analytically estimated by  Eq.~\eqref{eq:TF2}.
It can be seen from the plots that there is a good agreement between our estimated $T_F$ and the point where $S$ starts to deviate from its equilibrium distribution, meaning that our estimation catches the essence of the freeze out of the system.
The $S$ abundance appears to ``freeze out'' at $T_F$ because fusions of $S$ particles in (\ref{eq:proc-Qmin}) have stopped and---although the forward processes in (\ref{eq:proc-SS}) and (\ref{eq:proc-capture}) are still active---the $n_{S^\dagger}$ and $n_{Q<\Qmax}$ abundances are subdominant to $n_S$ and thus have little effect on a logarithmic scale.

Using Eq.~\eqref{eq:mu_analytic} to take into account the $T$ dependence of the chemical potential and equating $T_F = T_D$ with \eqref{eq:TD} and \eqref{eq:TF2}, 
we determine the parameter boundary to have Q-balls as the dominant component of dark matter in terms of $\eta$ and other NTS-related parameters, 
\begin{align}\label{eq:contour}
    \log\eta=\frac{m_S+m_{\Qmin}}{m_{\Qmin}}\log\left[\dfrac{2}{c_\gamma}\left(\dfrac{m_S}{2\pi \TFD}\right)^{\frac{3}{2}}\right]
    + \frac{m_S}{m_{\Qmin}} \log\left[\dfrac{\pi g^{1/2}_\ast c_\gamma \,\TFD^{1/2}}{\sqrt{90}\Qmax\Mpl\, (\sigma v_{\rm rel})_{\Qmin}}\left(\dfrac{4\pi^2 \TFD}{m_S\,m_{\Qmin}}\right)^{\frac{3}{2}}\right] ~.
\end{align}
To further understand how $\eta$ scales with $v$ or $\sigma_0$, 
we observe that $\TFD/v$ in~\eqref{eq:contour} can be treated approximately as a constant for this freeze-out system. Indeed, we have numerically tested and found that $T_F/v \approx 5~\text{to}~10$ is not sensitive to the other parameters in the system.
Because all the scales are proportional to one another, $T_F \propto v \propto m_S \propto m_{\Qmin}$,~\footnote{We expect the system we are discussing is of only one energy scale, \ie, the PT temperature should be roughly the same as $v$ or $\sigma_0$.} the right-hand side of \eqref{eq:contour} depends on $v$ only through the second term, which dominates the first term as $\Qmax$ becomes large.
Therefore, at large $\Qmax$, we expect $\eta$ to scale as 
\begin{align}\label{eq:eta-v-Qmax}
\eta\propto \left[\frac{v}{\Qmax\,\Mpl}\right]^{\frac{m_S}{m_{\Qmin}}}\,.
\end{align}
Note that $m_S/m_{\Qmin} \approx 1/\Qmin$,
giving the power-law dependence of the boundary.

\begin{figure}[t!]
    \centering
    \includegraphics[width=0.49\textwidth]{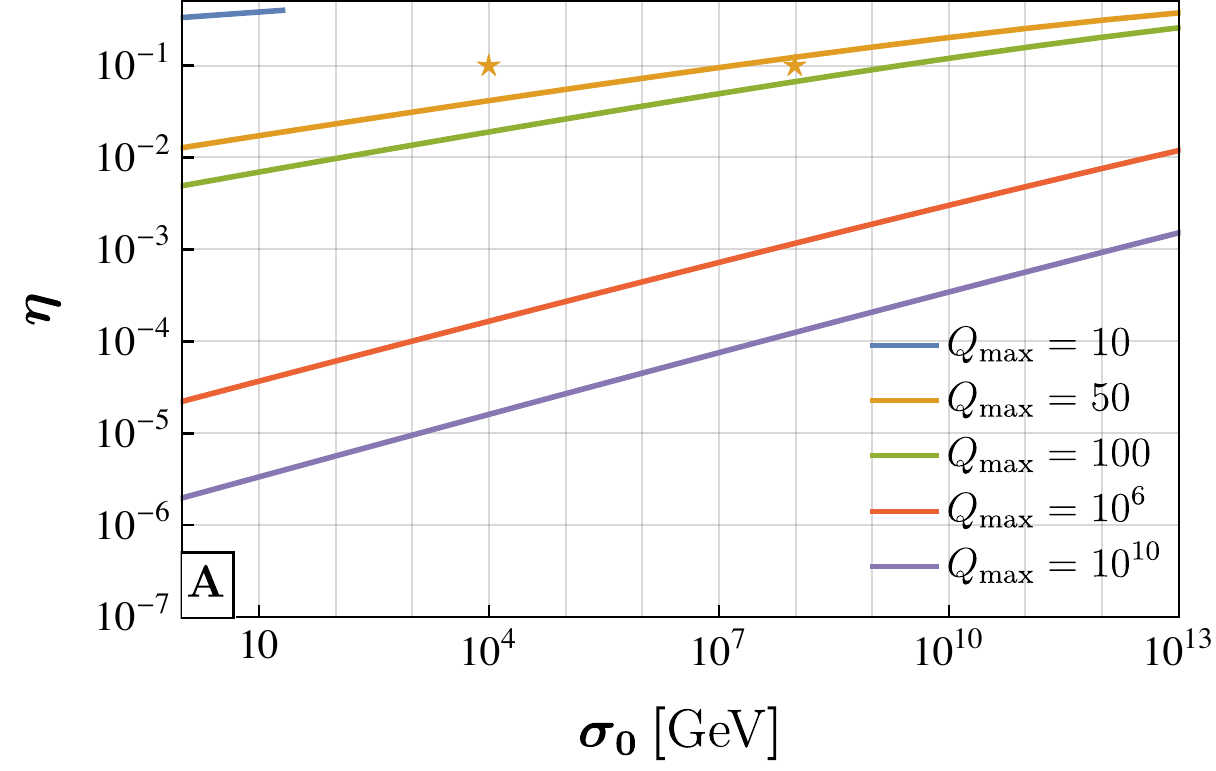}
    \includegraphics[width=0.49\textwidth]{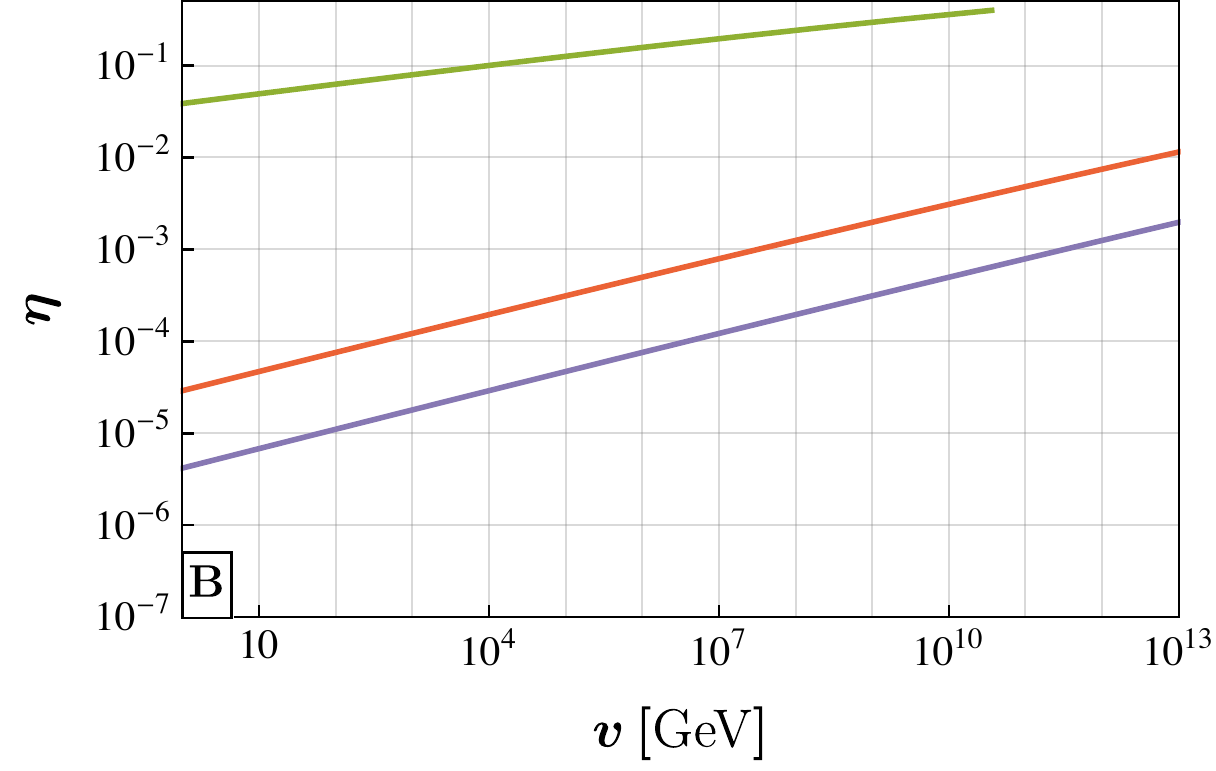}
    \caption{The contours of $T_D=T_F$ for Model A (left panel) and Model B (right panel) and various $\Qmax$.
    For all curves, $\Qmin=4$.
    Above each curve is the parameter space where solitosynthesis can happen with Q-balls dominant over free particles in charge abundance. For the left panel, the two star symbols correspond to the two benchmark points shown in Fig.~\ref{fig:Boltzmann}---one with ``efficient'' solitosynthesis and one without.
    Only the contours with $Q_{\rm max} = 100,~10^6,~\text{and}~10^{10}$ are shown in the right panel due to the plot range.
    }
    \label{fig:contours}
\end{figure}

In Fig.~\ref{fig:contours} we show the contours of $T_D=T_F$ for various $\Qmax$.
Above the contours are the regions where we expect efficient solitosynthesis to occur and NTSs to be the dominant component of charge in the dark sector.
It can be easily seen from the behavior of the contours that, as $\Qmax$ becomes large, the scaling of the contours indeed follows Eq.~\eqref{eq:eta-v-Qmax}.
Meanwhile, comparing the contours with the same $\Qmax$ in the two panels, we find that the boundaries of efficient solitosynthesis are similar for the two models when $\Qmax$ becomes large.
This is, again, because the second term in \eqref{eq:contour} determines the contour behaviors at large $\Qmax$ with a tiny model-dependent effect in the logarithm function.
Note that for the same set of $\Qmin$ and $\Qmax$, we predict here a smaller parameter space to have efficient solitosynthesis compared with Ref.~\cite{Griest:1989bq}, as their estimation renders a lower $T_F$.
But as we can use a much larger value of $\Qmax$ as justified in Sec.~\ref{sec:maximum-charge}, the available parameter space for efficient solitosynthesis turns out to be much larger than that in~\cite{Griest:1989bq}.

\begin{figure}[t!]
    \centering
    \includegraphics[width=0.48\textwidth]{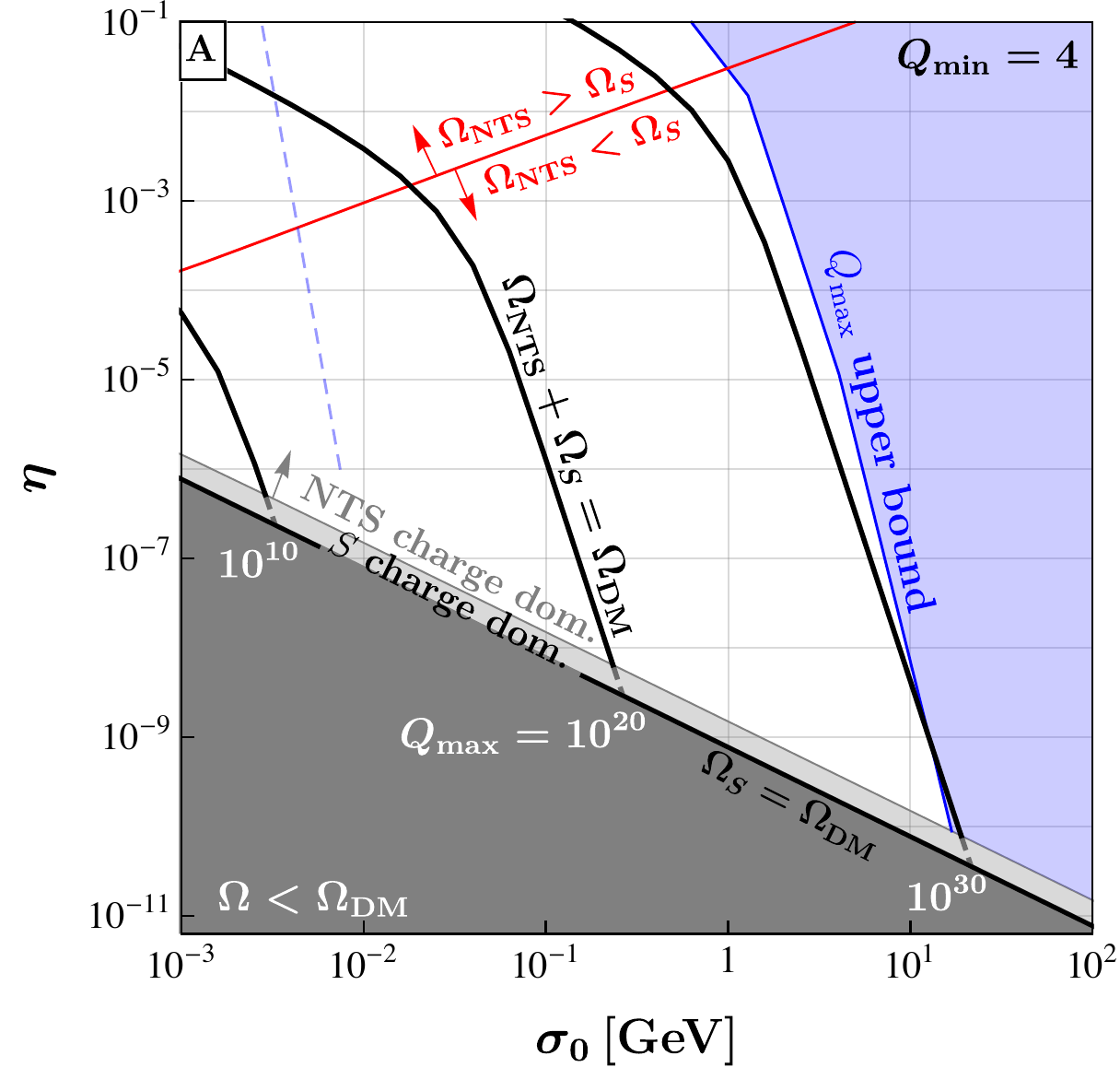}
    \includegraphics[width=0.47\textwidth]{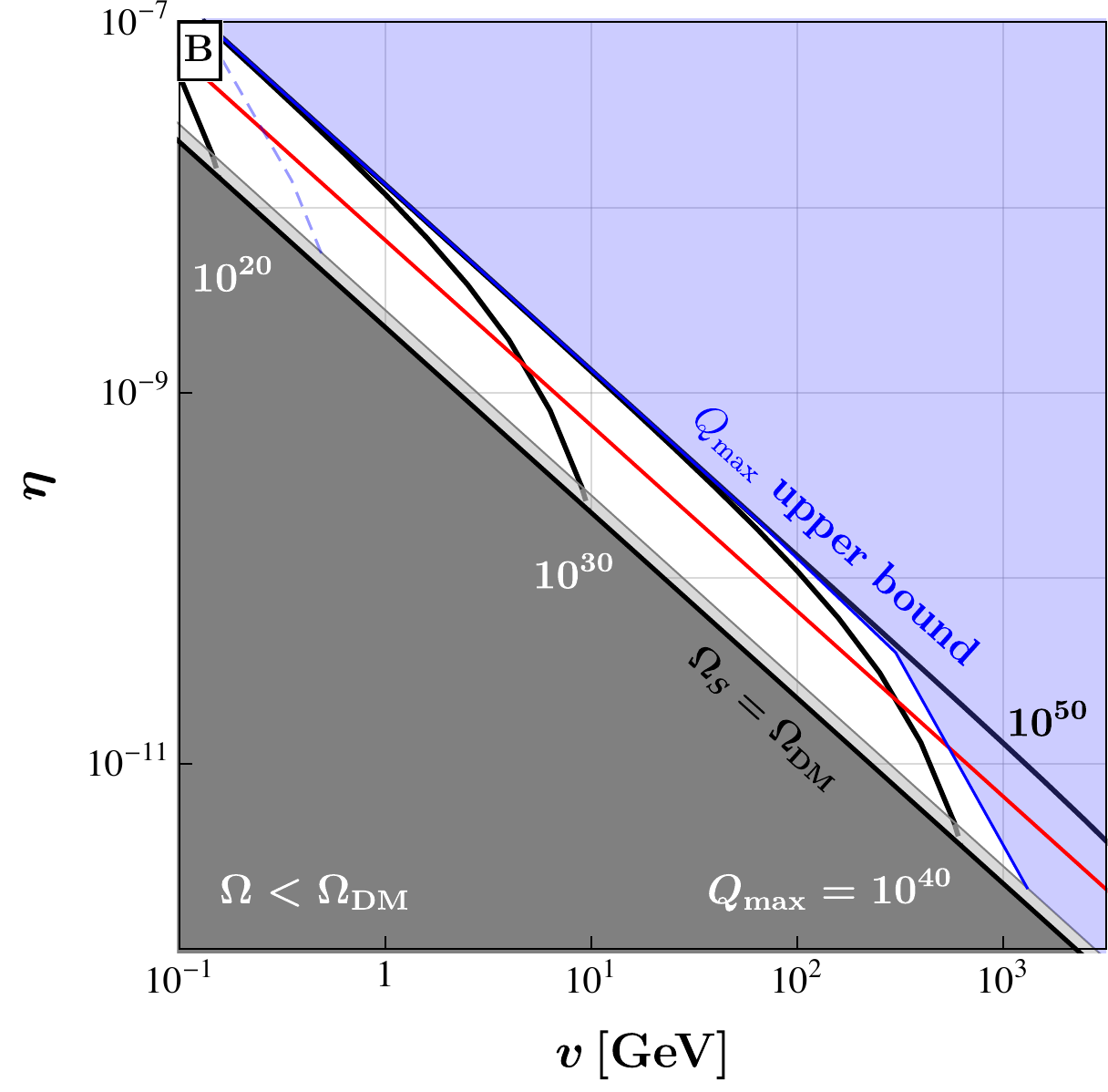}
    \caption{
    The parameter space to have the summation of Q-balls and free $S$ particles to explain the dark matter relic abundance $\Omega_{\rm DM} h^2=0.12$ after solitosynthesis for Model A (left) and B (right).
    The gray line separates the Q-ball and $S$-particle charge-dominated regions, marking the boundary of efficient solitosynthesis.
    For the region above the red line, the Q-ball energy abundance is higher than the free-particle energy abundance. Different black solid curves correspond to different choices of $\Qmax$, and they all converge to the line of $\Omega_S=\Omega_{\rm DM}$ as $\eta$ decreases (before they converge, the curves are plotted dashed in the light gray shaded region because the approximate formula in \eqref{eq:mu_analytic} breaks down here). 
    The blue shaded region is excluded by the upper bound on $\Qmax$ in \eqref{eq:Qmax_bound} (left) or \eqref{eq:Qmax_bound_2} (right) for $\Qmin=4$, where we have taken $T=\sigma_0$ or $v$.
    The thin blue dashed line is the same $\Qmax$ upper bound but for $\Qmin=7$, where for Model B we chose $\lc=6.2$ instead of 10.
    Otherwise, model parameters are taken as the benchmarks in Eq.~(\ref{eq:benchmark}). 
    }
    \label{fig:relic_abundance_SS}
\end{figure}

With the freeze-out temperature determined, we can now  derive the relic abundance of the dark sector (free particle+Q-balls), which combined with the boundary of solitosynthesis gives us the viable parameter space of the models.
In Fig.~\ref{fig:relic_abundance_SS} we show the parameter space where the relic abundance of the dark sector matches that of the dark matter, $\Omega_{\rm DM} h^2=0.12$~\cite{Planck:2018vyg}.
For each combination of $\eta$ and $\sigma_0$ or $v$ where solitosynthesis can happen, there is a unique $\Qmax$ to satisfy the relic abundance condition, denoted by the solid black lines (above these lines, the dark sector overcloses the Universe for fixed $\Qmax$). On the boundary for efficient solitosynthesis to occur (gray line separating NTS charge-dominated region from $S$ charge-dominated region), a smaller amount of charge asymmetry requires a larger $\Qmax$ and energy scale $v$ or $\sigma_0$.
On the other hand, $\Qmax$ is bounded from above by the equilibrium requirement as in \eqref{eq:Qmax_bound} and \eqref{eq:Qmax_bound_2} (excluding the blue shaded region), which is more stringent for larger $\sigma_0$ or $v$. 
As a result, large $\sigma_0$ or $v$ is excluded, while the exact boundary is model dependent due to the different $Q$ dependence in the cross section.

For each relic abundance curve of a certain $\Qmax$, there are two knees in the curve.
The one at larger $\eta$ (around the intersection with the red line) corresponds to the boundary between Q-ball and free particle dominance of the dark sector energy density, as explained at the end of Sec.~\ref{subsec:TD}.
Above this knee, the relic abundance is dominated by Q-balls, and thus can be calculated as
\begin{align}
\Omega_{\Qmax,0}=\dfrac{m_{\Qmax}n_{\Qmax,0}}{\rho_{c,0}}=\dfrac{m_{\Qmax}n_{\Qmax,T_F}}{\rho_{c,0}}\left(\dfrac{T_0}{T_F}\right)^3=\dfrac{(m_{\Qmax}/\Qmax)\eta\uu c_\gamma T^3_0}{\rho_{c,0}}\propto\dfrac{\eta\uu m_{\Qmax}}{\Qmax}\,,
\end{align}
where we have approximated $n_{\Qmax,T_F}\approx\eta\uu n_\gamma/\Qmax=\eta\uu c_\gamma T^3_F/\Qmax$,
and the subscript ``0" indicates the values today.
This explains why the relic abundance curves of different $\Qmax$ converge above this knee for Model B, as $m_{\Qmax} \propto \Qmax$ such that the $\Qmax$ dependence cancels out in the expression when $\Qmax$ is large.
For Model A, because $m_{\Qmax} \propto \Qmax^{3/4}$, different curves have separate behaviors. The second knee happens at a smaller $\eta$ where the relic abundance is completely determined by the free particle $S$, \ie, $\Omega_S=\Omega_{\rm DM}$.
The transition should be smooth, while in Fig.~\ref{fig:relic_abundance_SS} the transition is plotted dashed and sharp.
This is because we have used the approximation \eqref{eq:mu_analytic} in the calculation, which breaks down in the $S$ charge domination region, so the results are not reliable in the light gray shaded region. However, the general knee behavior is anticipated.~\footnote{For very small $\eta$ one should expect the situation to match the case of symmetric dark matter, which means that the $\Omega_S=\Omega_{\rm DM}$ curves should not continue as they are in Fig.~\ref{fig:relic_abundance_SS}, but have a unitary bound in $\sigma_0$ or $v$. This bound is beyond the range plotted and therefore is not shown here.}
In Fig.~\ref{fig:relic_abundance_SS} we have mainly used $\Qmin=4$. 
With a larger $\Qmin$, for a certain $\eta$ and $v/\sigma_0$ to keep having efficient solitosynthesis one needs to increase $\Qmax$ as well, as can be seen from Eq.~\eqref{eq:eta-v-Qmax}.
Constraints from the maximum attainable $\Qmax$ become more stringent and shift to smaller $v/\sigma_0$ in this case.
The thin blue dashed line shows the corresponding $\Qmax$ upper bound for $\Qmin=7$.~\footnote{The gray shaded region should shift for a different $\Qmin$, but we have checked that it is hardly noticeable for our new choice compared with the case of $\Qmin=4$.} 
The $\Omega_{\rm NTS}+\Omega_S=\Omega_{\rm DM}$ contours of fixed $\Qmax$ will also shift as $\Qmin$ becomes larger, so the thin blue dashed line should not be compared to the black lines in the figure. 
For Model A, the point where the blue solid line and the light gray line intersect corresponds to $\Qmax=6\times 10^{29}$, while for the blue dashed line of $\Qmin=7$ it is $\Qmax=5\times 10^{36}$, marking the increase of $\Qmax$ with respect to a larger $\Qmin$.
For Model B, the corresponding $\Qmax$ values are $1\times 10^{42}$ for $\Qmin=4$ and $2\times 10^{52}$ for $\Qmin=7$.
For both models, a larger $\Qmin$ prefers a smaller energy scale, which could be constrained by Big Bang nucleosynthesis observables depending on additional model details. 

Note, microlensing searches exclude compact objects with mass $m \gtrsim 10^{23}~\text{g}$ from making up all of dark matter \cite{Niikura:2017zjd,Smyth:2019whb,Macho:2000nvd,EROS-2:2006ryy,Wyrzykowski:2011tr,Griest:2013aaa,Oguri:2017ock}.
Our Model A is not constrained by this bound because the masses $m_Q \propto Q^{3/4}$ are too small. 
For Model B with a $\Qmax$ given in \eqref{eq:Qmax_bound_2}, microlensing requires $v\gtrsim 10^2$ GeV with benchmark (\ref{eq:benchmark}).

\section{Q-balls from a phase transition}
\label{sec:Q_ball_from_PT}

Q-balls can form from either an FOPT or SOPT~\cite{Frieman:1988ut,Griest:1989cb,Frieman:1989bx,Macpherson:1994wf,Hong:2020est,Bai:2018dxf,Ponton:2019hux,Bai:2021mzu} when the vacuum expectation value of $\sigma$ changes from $\sigma_0$ to $\sigma_-$ for Model A and that of $\phi$ changes from zero to $v$ for Model B. When regions of false vacuum contain a charge $Q >\Qmin$, it can be energetically preferable to remain in the false vacuum and lead to Q-balls. This could come about either due to some initial asymmetry in the $S$ sector, or simply due to statistical fluctuations in the difference of $S$ particles and antiparticles within the relevant volumes \cite{Griest:1989cb}.
Once formed, these Q-balls could serve as the initial seeds for solitosynthesis, as discussed in the previous section.
But if the chemical equilibrium between free $S$ particles and the Q-balls cannot be reached, this initial population could remain relatively unchanged since the end of the PT.

The initial conditions after a PT can be modified in two ways, possibly simultaneously: either (a) by  the building up of Q-balls starting from fusion of free particles to form $(\Qmin)$ or (b) from the evolution of the Q-balls formed during the PT. 
We are interested in the possibility that the initial conditions are not modified.
There are various reasons for (a) not to occur.
For example, it is plausible that the PT temperature $T_f$ is lower than $T_F$, such that there is not enough time for the chemical equilibrium to be established.
Note that $T_F$ is usually smaller than $\sigma_0$ or $v$ by an $\mc{O}(1)$ number, so the potential might not have to be very fine-tuned to achieve a lower PT temperature. 

Regarding (b), a sufficient condition for no evolution of the PT-produced Q-balls is that the number of free particles in the true vacuum regions is highly suppressed after the PT.
This can occur if (i) the free particles remain inside the false vacuum bubbles due to energetics and bubble wall dynamics, (ii) the free particles are not thermally produced in the reverse of (\ref{eq:proc-SS}), and (iii) the thermal bath temperature is low enough that it cannot dislodge particles from Q-balls in the reverse of (\ref{eq:proc-capture}) or (\ref{eq:proc-Qsbar}). All three conditions can be true provided $m_S/T_f$ is large enough, perhaps requiring special model engineering. For (i), in an FOPT, the proportion of particles trapped in the false vacuum can be order unity for $m_S/T_f \gtrsim \mathcal{O}(10)$, depending on the bubble wall velocity \cite{Hong:2020est}. In an SOPT particles are initially distributed randomly in true and false vacuum pockets, but then may be expected to rearrange to favor the false vacuum pockets. Because they are heavier inside true vacuum pockets, the probability to remain there should be suppressed by a Boltzmann factor $\sim e^{-m_S/T}$. There may also be additional bubble wall dynamics similar to the FOPT as false vacuum pockets shrink~\cite{Frieman:1988ut}. For (ii), similar to \cite{Baker:2019ndr}, we require $(\sigma v_\text{rel}) n_S^{\rm eq} \lesssim H$ using (\ref{eq:SS-xsec}) and (\ref{eq:number_density}) with $\mu_S=0$ at $T=T_f$, giving $m_S/T_f \gtrsim 31 + (3/2)\log [m_S/(31 T_f)] + \log (T_f\cdot\text{TeV}/v^2)$.
Finally, for (iii), because the binding energy per $S$ particle in a large-charge Q-ball is generically of order the free particle mass in Model B [barring a fine tuning with $(\ls \lphi)^{1/4}$ very close to $m_S/v$] and could be significantly larger in Model A, it is energetically unlikely for $S$ or $S^\dagger$ particles to be kicked out of Q-balls. 
Specifically, detailed balance allows the estimation that these processes are irrelevant when $(\sigma v_\text{rel})_{\langle Q \rangle} n_S^{\rm eq}<H$ with $\mu_S=0$, giving $m_S/T_f \gtrsim 50 + (3/2)\log [m_S/(50\,T_f)] + \log (T_f\cdot\text{TeV}/v^2) + (1/2)\log(\langle Q \rangle/10^{10})$ for Model A and $m_S/T_f \gtrsim 53 + (3/2)\log [m_S/(53\,T_f)] + \log (T_f\cdot\text{TeV}/v^2) + (2/3)\log(\langle Q \rangle/10^{10})$ for Model B, where $\langle Q \rangle$ is the typical charge produced from the PT.
In this case, it is unlikely for chemical equilibrium to be achieved following the PT, and the solitosynthesis story in Sec.~\ref{sec:solitosynthesis} needs not apply. If these conditions do not hold, we must carefully consider how the charge and abundance of Q-balls and free particles evolve.

To estimate the typical Q-ball charge immediately following a phase transition, we begin with the number density of $S$ and $S^\dagger$ particles in a given Hubble volume near the temperature $T_f$: 
$n_S = (2 \zeta(3)/\pi^2) T_f^3$,
where $T_f$ refers to the bubble nucleation temperature $T_n$ (when the true vacuum occupies $1-e^{-1}$ of the total volume) for an FOPT or the Ginzburg temperature $T_G$ (the temperature where thermal fluctuations between the true and false vacua freeze out) for an SOPT. These $S$ particles will be divided up into a number of potential pockets for Q-ball formation. 

For an FOPT, the number of potentially formed Q-balls approximately equals the number of bubble nucleation sites. This number is determined by the temperature-dependent bounce action for the field to transition from the false to the true vacuum
(see Appendices~\ref{appendix:bounce_action} and \ref{appendix:nucleation_sites} for more detailed discussion). Once bubbles have nucleated, $S$ particles will be ``snowplowed'' by the bubble walls owing to their smaller mass inside the false vacuum. Where bubble walls meet, $S$ particles can collect and form Q-balls.
Parametrizing the bounce action of the FOPT as $S_3/T = a/\epsc^2$, with $\epsc = (T_c-T)/T_c$ and $T_c$ is the temperature when the two vacua are degenerate, the number density of Q-balls formed at the bubble nucleation temperature $T_n$ is (see Appendix~\ref{appendix:nucleation_sites}) 
\begin{equation}
n_\text{Q-ball}(T_n) \sim n_\text{nuc} \approx (4 \pi v_\text{sh}^3 a^{1/2})^{-1} H_n^3 \left(\log \left[\frac{v_\text{sh}^3\, \epsn^9\, T_n^4}{8 \sqrt{2\pi}\, a^{5/2} \, H_n^4}\right]\right)^{3/2} \, , \; \; \; \; \text{(FOPT)} .
\label{eq:nQball_FOPT}
\end{equation}
Here, subscript $n$ denotes quantities evaluated at $T=T_n$, and we have taken $\epsn = \epsc|_{T=T_n} \lesssim 1$.

For an SOPT, the number of Q-ball-forming sites depends on the correlation length $\xi$ of the PT and the probability for each correlated region to be in the false vacuum \cite{Frieman:1988ut}. The latter depends on the energy difference between the false and true vacua at the Ginzburg temperature $T_G$. The probability ratio is $p_\text{false}/p_\text{true} \sim \exp [-\Delta V(T_G) \, (2\xi)^3 / T_G]$. The correlation length also depends on the Ginzburg temperature as $\xi \simeq (\lphi T_G)^{-1}$ for Model B (and replacing $\lphi$ by $\lambda$ for Model A), where $T_G \simeq \lphi^{-1/2} v$~\cite{Frieman:1988ut} and $\Delta V(T_G) \sim \lphi v^4 /4$. Thus, $p_\text{false}/p_\text{true} \sim e^{-2}$. 
This gives the number density of potentially Q-ball-forming correlated regions 
\begin{align}
n_\text{Q-ball} (T_G) \sim 
\frac{1}{1+p_\text{true}/p_\text{false}} \xi^{-3} \sim 10^{-1} \lphi^{3/2} v^3 \, , \; \; \; \; \text{(SOPT)}\,.
\end{align}
Notice that this is generally orders of magnitude larger than the number density from an FOPT because $v \gg H_n \sim v^2/\Mpl$. Thus, an SOPT will produce more numerous but smaller-charged Q-balls.

The number of $S$ particles and antiparticles within the proto-Q-ball is $N_S^\text{Q-ball} \sim p_\text{in} \, n_S/ n_\text{Q-ball}$. The factor $p_\text{in}$ accounts for the probability for each of the $S$ particles to remain inside the false vacuum regions and form Q-balls. 
The typical Q-ball charge will be the greater of the asymmetric or statistical fluctuation components: $\langle Q \rangle \sim \max \left[\eta N_S^\text{Q-ball}, \, (N_S^\text{Q-ball})^{1/2} \right]$, where $\eta \approx |n_S - n_{S^\dagger}| / (n_S + n_{S^\dagger})$. 
\footnote{The approximation is exact when $S$ particles are relativistic so $n_S+n_{S^\dagger} = n_\gamma$. If more exact results are required, simply define $\eta$ using $n_S+n_{S^\dagger}$ instead of $n_\gamma$ in the denominator for the purposes of this section.} Notice that if the asymmetric component (first term) dominates $\langle Q \rangle$, then most Q-balls will have same-sign charge, whereas if the statistical fluctuations (second term) dominate, then both positively and negatively charged Q-balls result in equal proportion.

The Q-ball abundance can also be calculated as $Y_\text{Q-ball} \sim p_{Q>\Qmin} n_\text{Q-ball}\,s^{-1}$.
The factor $p_{Q>\Qmin}$ accounts for the probability for each proto-Q-ball to have large enough charge to be stable. Often, $\langle Q \rangle \gg \Qmin \sim \mathcal{O}(1\text{ to } 10^3)$, so this factor can be $\mathcal{O}(1)$. The factor $s=(2\pi^2/45) g_{*S} T^3$ is the entropy density with $g_{*S} \sim 100$. This can be compared to the observed dark matter abundance $Y_\text{DM} = (3.6 \times 10^{-10}) (\GeV/m_Q)$. 
For an FOPT and the mass spectrum of $m_Q = 5.15 \sigma_0 \lambda^{1/4} \langle Q\rangle^{3/4}$, 
\begin{align}
\frac{Y_\text{Q-ball}}{Y_\text{DM}} \sim & (1.3 \times 10^{-5}) p_{Q>\Qmin} g_{*s}^{-1} \lambda^{1/4} \left(\frac{\sigma_0}{\GeV}\right)^{7/4} a^{-1/8} 
\\
& \times \max \left( \eta \,p_\text{in}^{3/4} g_*^{3/8} l^{3/8} v_\text{sh}^{-3/4} \, , \, \, (6.4 \times 10^{-23}) \left(\frac{\sigma_0}{\GeV}\right)^{9/8} \frac{p_\text{in}^{3/8} g_*^{15/16} l^{15/16}}{a^{3/16} v_\text{sh}^{15/8}}  \right) ~, 
\nonumber
\end{align}
while for the mass spectrum $m_Q = \langle Q \rangle v (\lphi \ls)^{1/4}$,
\begin{align}
\frac{Y_\text{Q-ball}}{Y_\text{DM}} \sim & (1.5\times 10^{9}) p_{Q>\Qmin} g_{*s}^{-1} (\lphi \ls)^{1/4}\frac{v}{\GeV} 
\\ 
& \times \max \left(\eta \, p_\text{in} \, , \, \, (2.6 \times 10^{-30}) p_\text{in}^{1/2} \left(\frac{v}{\GeV}\right)^{3/2} \left(\frac{g_*^3\uu  l^3}{v_\text{sh}^6 a}\right)^{1/4} \right) \, ,
\nonumber
\end{align}
where $l\equiv \log \left[v_\text{sh}^3\, \epsn^9\, T_n^4/(8 \sqrt{2\pi}\, a^{5/2} \, H_n^4)\right]$. Meanwhile, for an SOPT and the mass spectrum of $m_Q = 5.15 \sigma_0 \lambda^{1/4} \langle Q\rangle^{3/4}$, 
\begin{align}
\frac{Y_\text{Q-ball}}{Y_\text{DM}} \sim & (1.1 \times 10^{10}) p_{Q>\Qmin} g_{*s}^{-1} \lambda \frac{\sigma_0}{\GeV} \max \left(\frac{\eta\, p_\text{in}^{3/4}}{(1+p_\text{true}/p_\text{false})^{1/4}} \, , \, \, \frac{1.7 \lambda^{9/8} p_\text{in}^{3/8}}{(1+p_\text{true}/p_\text{false})^{5/8}} \right)~,
\end{align}
while for the spectrum of $m_Q = \langle Q \rangle v (\lphi \ls)^{1/4}$,
\begin{align}
\frac{Y_\text{Q-ball}}{Y_\text{DM}} \sim & (1.5 \times 10^{9}) p_{Q>\Qmin} g_{*s}^{-1} (\lphi \ls)^{1/4}\frac{v}{\GeV} \max \left(\eta\,  p_\text{in} \, , \, \, 2 \lphi^{3/2} \sqrt{\frac{p_\text{in}}{1+p_\text{true}/p_\text{false}}} \right) ~.
\end{align}

\begin{figure}[t]
\centering
    \includegraphics[width=0.48\textwidth]{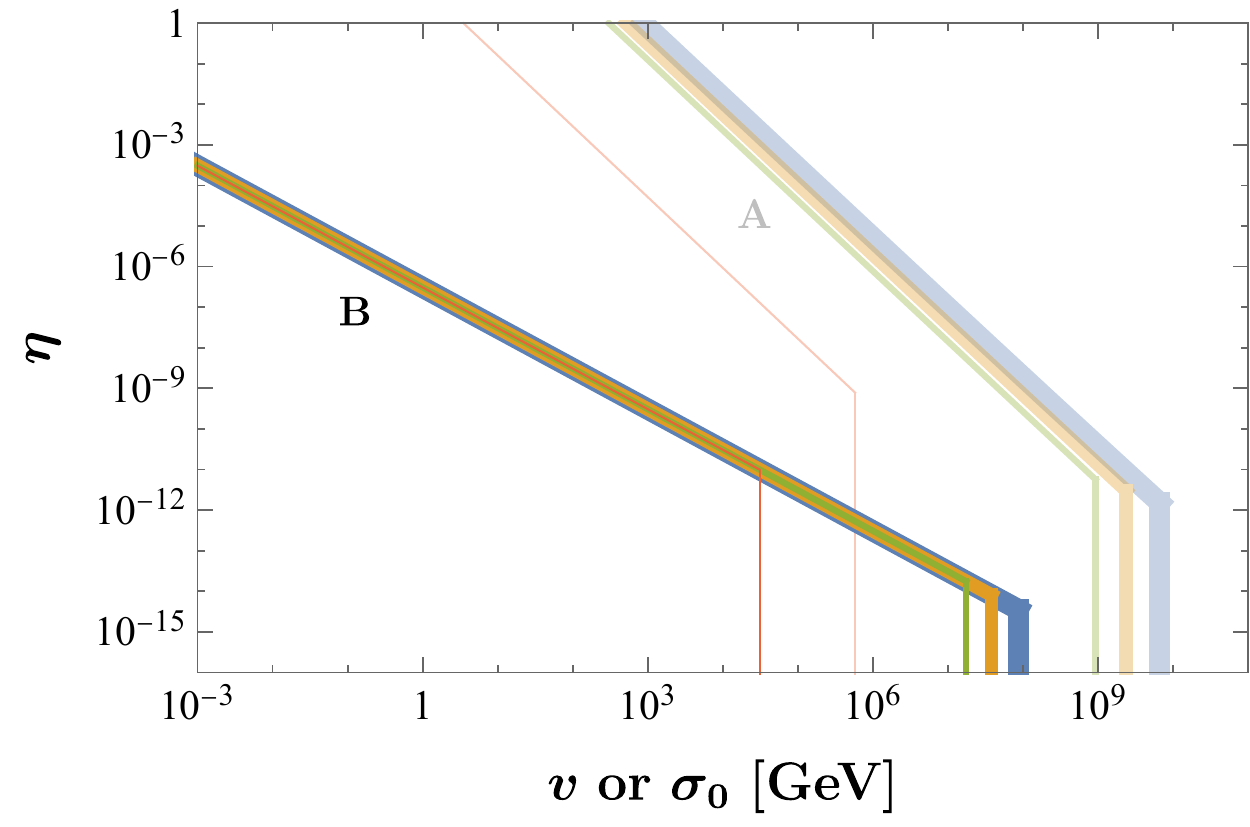} \hspace{4mm}
    \includegraphics[width=0.48\textwidth]{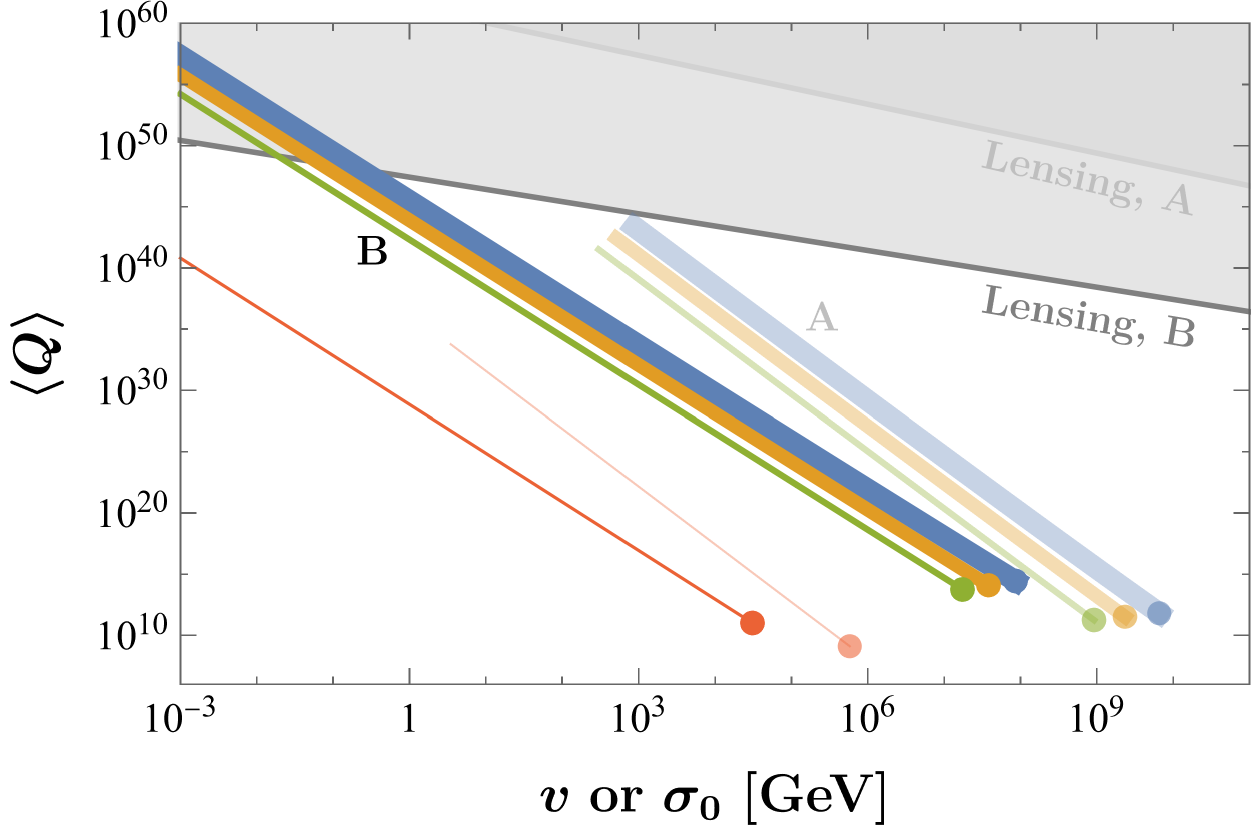}
    \caption{{\it Left panel:} The values of $\eta$ and $v$ or $\sigma_0$ that give the right abundance of Q-balls for them to make up all dark matter from an FOPT without solitosynthesis. {\it Right panel:} The corresponding average charges of the Q-balls produced making up all dark matter. The points at the bottom right of each curve represent the values consistent with $\eta = 0$ and the statistical fluctuations in the symmetric component of the $S$ particles dominate, {\it i.e.}, where the curves become vertical in the left panel. 
    In both panels, Models A and B are plotted in lighter and darker colors, respectively, and the constraint $\eta<1$ has been imposed. 
    Here, $T_n=v$ or $\sigma_0$, $v_\text{sh}=1$, $\epsilon_n=0.01$, $p_\text{in}=1$, and the curves correspond to---from thickest to thinnest and right to left---the values $a=100$ (blue), $10^{-1}$ (yellow), $10^{-4}$ (green), and $10^{-30}$ (red). Microlensing constraints on $m_Q>10^{23}~\text{g}$ are shown in shaded gray \cite{Niikura:2017zjd,Smyth:2019whb,Macho:2000nvd,EROS-2:2006ryy,Wyrzykowski:2011tr,Griest:2013aaa,Oguri:2017ock}.
    }
    \label{fig:1OPT}
\end{figure}

\begin{figure}[t]
\centering
    \includegraphics[width=0.48\textwidth]{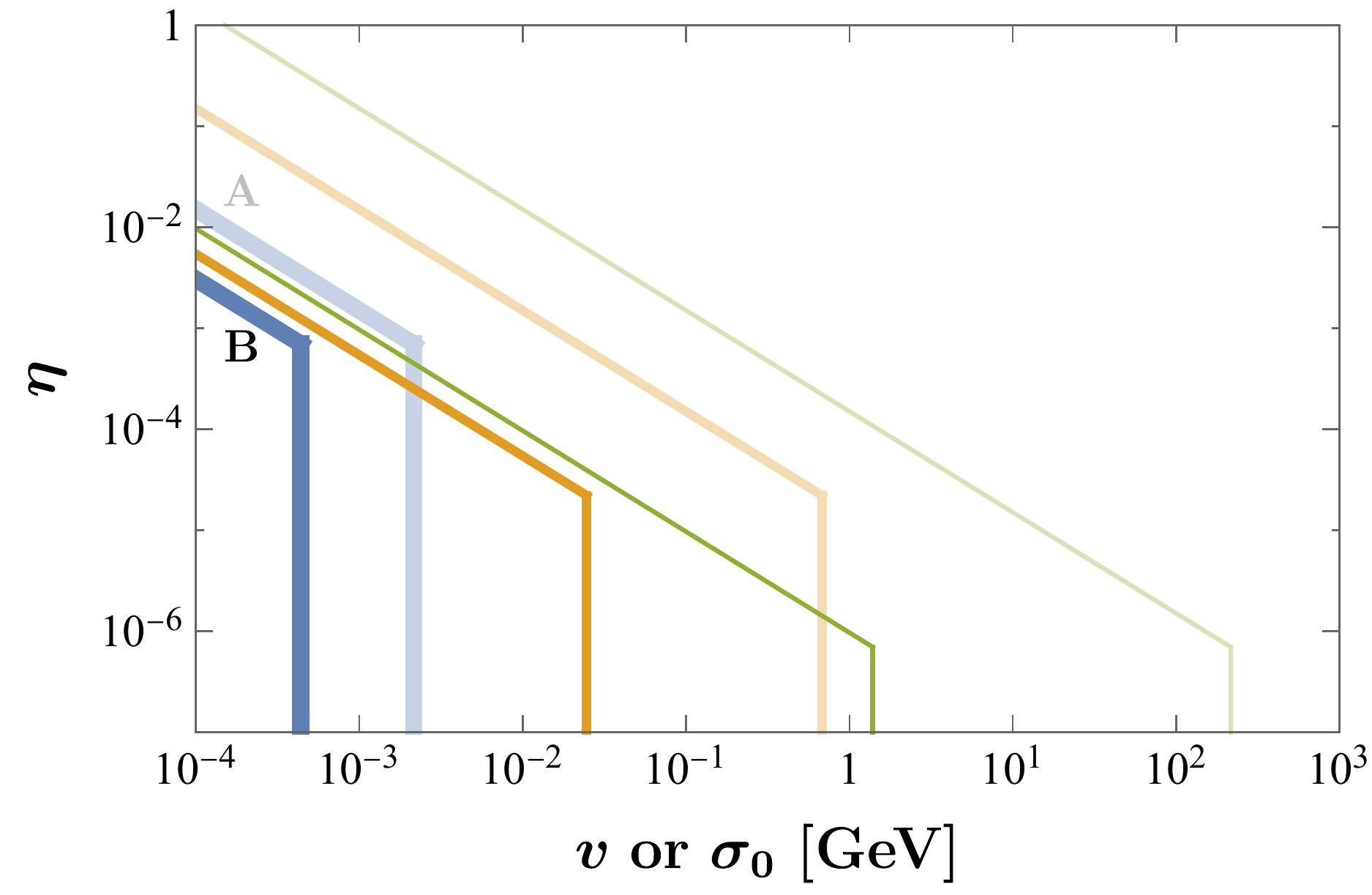} \hspace{4mm}
    \includegraphics[width=0.48\textwidth]{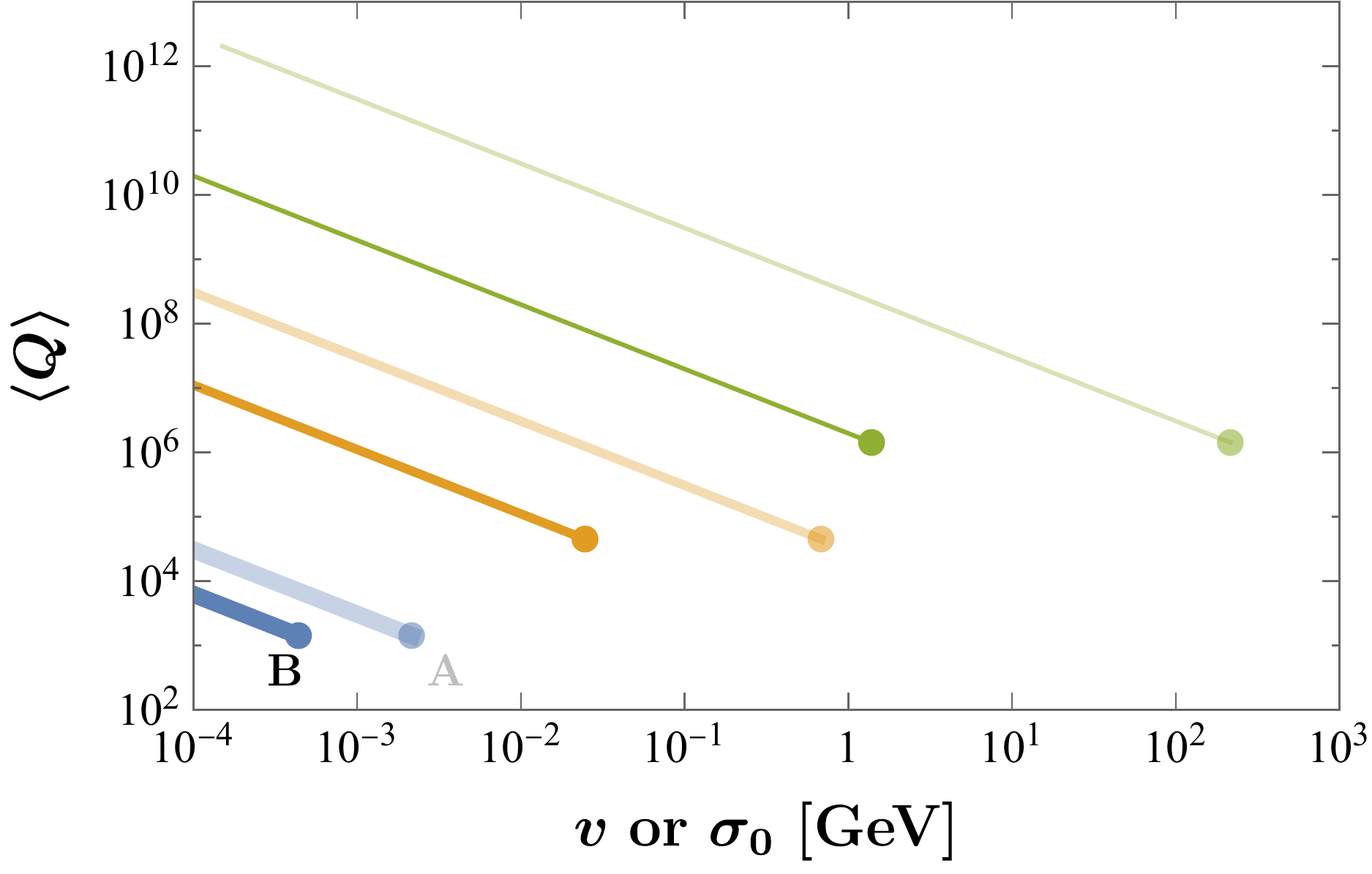}
    \caption{Similar to Fig.~\ref{fig:1OPT}, but for an SOPT. Here, $T_G= \lphi^{-1/2} v$, $p_\text{in}=1$, and $p_\text{false}/p_\text{true}=e^{-2}$.
    The two models considered are $m_Q=5.15 \lphi^{1/4} \sigma_0 Q^{3/4}$ (A, lighter shading) and $m_Q=(\lphi \ls)^{1/4} Q v$ with $\ls=0.2$ (B, darker shading). Note that here Model A and B share the same $\lphi$. The curves correspond to the values $\lphi=10^{-2}$ (blue thick), $10^{-3}$ (yellow medium), and $10^{-4}$ (green thin). 
    Curves with $\lphi \gtrsim 10^{-1}$ would require $T_G,m_S < \text{MeV}$, which is disallowed by the constraints on radiation degrees of freedom during Big Bang nucleosynthesis.
    }
    \label{fig:2OPT}
\end{figure}

In Figs.~\ref{fig:1OPT} and \ref{fig:2OPT} (corresponding to FOPTs and SOPTs, respectively), we show contours of parameters that give the proper abundance of Q-balls in the left panels, assuming $p_\text{in}=p_{Q>\Qmin}=1$. The right panels give the corresponding typical charges $\langle Q \rangle$ for those parameters (Model A curves stop at large $\langle Q\rangle$ when $\eta=1$). At larger $\eta$, the asymmetric component dominates the Q-ball properties. However, at smaller $\eta$ where the contours become vertical in the left panels and dots in the right panels, the symmetric component dominates, equivalent to the $\eta \to 0$ limit. For FOPTs, $a \lesssim 100$ for the expansion in $\epsc$ to hold, but there is in principle no lower bound. Note that because the correlation lengths for SOPTs tend to be much smaller than the bubble separation lengths of FOPTs, SOPTs tend to form a higher number density of Q-balls with smaller charges and therefore prefer lower energy scales compared to FOPTs. Additionally, Model A prefers higher charges and energy scales than Model B because Model A contains lighter Q-ball states with $m_Q \propto Q^{3/4}$.

\section{Discussion and conclusions}
\label{sec:conclusion}

Before concluding, we present some useful benchmarks for Q-ball properties formed from solitosynthesis or PTs. A plausible value for the amount of charge asymmetry is $\eta\sim 10^{-10}$, similar to the observed baryon asymmetry in the Standard Model sector. In the case of efficient solitosynthesis and for the $\Qmin=4$ benchmark presented in Fig.~\ref{fig:relic_abundance_SS}, the $\Qmax$ upper bound could give a reasonable estimate for the typical NTS charge $\langle Q \rangle$. For the PT formation, on the other hand, it is most interesting to note that $\eta=0$ gives viable parameter space. Using these assumptions, some benchmark values are given in Table~\ref{tab:benchmarks} for solitosynthesis, an FOPT, and an SOPT for both Models A and B with Q-balls making up all of dark matter (except solitosynthesis of Model A, in which $S$ particles dominate the energy density but Q-balls dominate the global charge for the parameter choice $\eta=10^{-10}$, see Fig.~\ref{fig:relic_abundance_SS}). Comparing these three examples, solitosynthesis produces Q-balls with the largest charges, masses, and radii. FOPTs prefer the largest values of $v$ or $\sigma_0$, giving Q-balls the smallest radii---despite not having the smallest charges---as well as macroscopic masses. SOPTs prefer the smallest values of the charges and of $v$ or $\sigma_0$, giving Q-balls microscopic masses. Containing less compact Q-ball states, Model B tends to produce larger-radius Q-balls than Model A. Much larger masses and radii are accessible with larger $\eta$, which is also demonstrated in Table~\ref{tab:benchmarks} by choosing a benchmark with $v$ near the MeV scale for solitosynthesis or an FOPT in Model B. Note that microlensing searches exclude compact objects with mass $m \gtrsim 10^{23}~\text{g}$ from making up all of dark matter \cite{Niikura:2017zjd,Smyth:2019whb,Macho:2000nvd,EROS-2:2006ryy,Wyrzykowski:2011tr,Griest:2013aaa,Oguri:2017ock}, and there are possibilities to probe down to much lower masses in the future \cite{Katz:2018zrn,Bai:2018bej,Jung:2019fcs}. Indeed, some of the points in Table~\ref{tab:benchmarks} and Figs.~\ref{fig:relic_abundance_SS} and \ref{fig:1OPT} are already constrained. Other detection strategies at still lighter mass \cite{Carney:2022gse}, such as multiple-scatter searches \cite{Bramante:2018qbc,DEAPCollaboration:2021raj}, rely on model-dependent couplings to the Standard Model \cite{Ponton:2019hux,Bai:2019ogh,Bai:2021mzu}, although NTSs with masses near the Planck scale could eventually be discovered through gravitational interactions alone \cite{Carney:2019pza,Windchime:2022whs}.

\begin{table}[t]
  \renewcommand{\arraystretch}{1.3}
    \addtolength{\tabcolsep}{3pt} 
    \centering
    \begin{tabular}{l| l|l||l | l| l |l}
    \hline \hline
         Mechanism & Model & $\eta$ & $m_Q$ (g) & $R_Q$ (m) & $\langle Q \rangle$ & $\sigma_0$  or $v$ 
         (GeV) \\ \hline
         \multirow{3}{*}{Solitosynthesis} & A & $10^{-10}$ & 3 & $3 \times 10^{-10}$ & $6 \times 10^{29}$ & 10 \\
         & B & $10^{-10}$ & $5\times 10^{22}$ & $2\times 10^{-3}$ & $1 \times 10^{45}$ & $1\times 10^2$ \\
         & B & $10^{-6}$ & $6\times 10^{30}$ & $3\times 10^5$ & $1\times 10^{57}$ & $1 \times 10^{-2}$ \\\hline
         \multirow{3}{*}{FOPT} & A & 0 & $9\times 10^{-6}$ & $5\times 10^{-23}$ & $3 \times 10^{11}$ & $2\times 10^9$ \\
         & B & 0 & $2\times 10^{-3}$ & $4 \times 10^{-19}$ & $1 \times 10^{14}$ & $4 \times 10^7$ \\ 
         & B & $10^{-4}$ & $8 \times 10^{26}$ & $1 \times 10^{5}$ & $7 \times 10^{53}$ & $3 \times 10^{-3}$ \\\hline
         \multirow{2}{*}{SOPT} & A & 0 & $2 \times 10^{-20}$ & $3\times 10^{-15}$ & $5 \times 10^4$ & $7 \times 10^{-1}$ \\
         & B & 0 & $1 \times 10^{-20}$ & $5 \times 10^{-13}$ & $5 \times 10^4$ & $2 \times 10^{-2}$ \\
         \hline \hline
    \end{tabular}
    \caption{
    Benchmark Q-ball properties in Models A and B as produced from solitosynthesis  or a PT making up all of dark matter (except solitosynthesis of Model A, see text). For the FOPT, $a=10^{-1}$ is assumed. Model parameters are chosen as in (\ref{eq:benchmark}) with $\Qmin=4$, except the SOPT where $\lambda=\lphi=10^{-3}$ is used. The benchmark points with the largest masses for solitosynthesis and FOPT were chosen to demonstrate the full range of possible properties, but they are excluded by microlensing searches from making up all of dark matter \cite{Niikura:2017zjd,Smyth:2019whb,Macho:2000nvd,EROS-2:2006ryy,Wyrzykowski:2011tr,Griest:2013aaa,Oguri:2017ock}.
    }
    \label{tab:benchmarks}
\end{table}

As Q-balls are usually massive macroscopic objects, it is natural to inspect the possibility that they collapse into black holes~\cite{Lee:1986ts,Friedberg:1986tq,Lee:1991ax}. We have found that the solitons from both solitosynthesis and PTs do not collapse into a black hole.
A simple criterion for the collapse is to compare the Q-ball radius with the Schwarzschild radius of a black hole with the same mass, \ie, requiring $R_Q \gtrsim G\,m_Q/2$. 
This defines a critical charge $Q_\text{BH}$ above which Q-balls collapse into black holes.
For solitosynthesis, one may compare this with the upper bounds given in \eqref{eq:Qmax_bound} and \eqref{eq:Qmax_bound_2}.
For $Q_\text{BH} < \Qmax$, for Model A, this leads to $n^2_S(T)/T^4\lesssim 0.8g^{1/2}_\ast/G$,
which sets $\sigma_0\lesssim 1.9\times 10^{21}$ GeV when taking $T=\sigma_0$ and $g_\ast=100$. Thus, Q-balls will not form for sub-Planckian values of $\sigma_0$.
For Model B, the corresponding bound [again, taking $m_Q=(\lambda_S\lambda_\phi)^{1/4}v\uu Q$] is $n^2_S(T)/(v^2 T^4) \lesssim 64 g_\ast \lambda^{1/2}_\phi/(3\lambda^{1/2}_S)$,
which sets a bound on the model parameters instead to be $\lambda^2_{\phi S}K^2_2(\lambda^{1/2}_{\phi S}/2)<4096\pi^4 g_\ast \lambda^{1/2}_\phi/(3\lambda^{1/2}_S)$.
The left-hand side of this inequality has a maximum value of 64 at $\lambda_{\phi S}=0$, and therefore the constraint holds easily unless $\lambda_\phi$ is as small as $10^{-10}$.

An assumption we made during the analysis of solitosynthesis is that the evolution process results in only a single thermalized system of free particles and Q-balls of charges from $\Qmin$ to $\Qmax$.
There is another possibility where the Q-balls formed from PTs are so large that they can only discharge to a charge of $Q_{\rm low}$ larger than $\Qmax$, the maximum charge available through fusion.
In other words, there is a gap in the Q-ball charge spectrum, and the system has two subsets that are not thermalized with each other.
Either subset could end up dominating the charge/energy density. Note that such a situation invalidates the chemical potential relationships \eqref{eq:mu} as well as the necessary ingredients to derive the Boltzmann equation \eqref{eq:boltz-NTS}.
In this case the analysis is more complicated, beyond the scope of this work. However, given the typical charges shown in Table~\ref{tab:benchmarks}, it may be more likely that $Q_\text{low}<\Qmax$ and there are not two separate subpopulations.

Neither of the two models we discussed has an intrinsic upper bound on the Q-ball charge, whose existence can change soliton evolution calculations in this work.
Such an upper bound does exist in some models, potentially because large Q-balls could destabilize the false vacuum~\cite{Kusenko:1997hj} or because a repulsive interaction within the large Q-balls is too strong for Q-balls to exist~\cite{Lee:1988ag}.
If solitosynthesis pushes the charges to beyond this upper bound, there may be unique signatures coming from, \eg, the induced phase transition \cite{Croon:2019rqu} or collapse of these large Q-balls in these respective examples. 

To conclude, in this work we study the cosmic evolution of Q-balls, considering both the scenarios with and without solitosynthesis.
In the case that solitosynthesis is efficient, we estimate which species dominates the total charge or energy of the free-particle--NTS system, and examine both analytically and numerically the evolution of the system through a set of coupled Boltzmann equations.
We then derive the parameter space where solitosynthesis is efficient such that Q-balls dominate the charges and/or energy density of the dark sector, and meanwhile the dark sector can make up all the dark matter.
Our calculations refine previous estimations on the maximum attainable charge in the system and the freeze-out temperature of NTS number density, and thus we find a much larger parameter space for efficient solitosynthesis.
Without solitosynthesis, Q-balls do not appreciably evolve after the PT. 
We discuss the possible reasons for this scenario and examine how the PT parameters determine the typical charge, mass, and abundance of the Q-balls.
Our work is restricted within scalar theories, while the discussions can be generalized to fermionic macroscopic states as well. The results of this work demonstrate that NTSs can be copiously produced in the early universe and serve as one type of macroscopic dark matter.

\subsubsection*{Acknowledgements}
The work of YB is supported by the U.S. Department of Energy under the contract DE-SC-0017647. The work of SL is supported in part by Israel Science Foundation under Grant No. 1302/19, and also by the Area of Excellence (AoE) under the Grant No. AoE/P-404/18-3 issued by the Research Grants Council of Hong Kong S.A.R. The work of NO is supported by the Arthur B. McDonald Canadian Astroparticle Physics Research Institute. We are grateful to the Munich Institute for Astro- and Particle Physics (MIAPP), which is funded by the Deutsche Forschungsgemeinschaft (DFG, German Research Foundation) under Germany's Excellence Strategy-EXC-2094-390783311, and the Mainz Institute for Theoretical Physics (MITP) of the Cluster of Excellence PRISMA+ (Project ID 39083149) for their hospitality and partial support during the completion of this work.

\appendix

\section{Analytic approximations for small $Q$}
\label{appedix:smallQ}

For Model B in (\ref{eq:V}), it has in the past been pointed out that the ansatz $s(r) \approx s_0 \sin (\omega r)/(\omega r)$ for $r<\pi/\omega$ provides a reasonable approximation to the solution for the field equations of motion \cite{Ponton:2019hux,Bai:2021mzu}. Indeed, this is the approximate solution when $f=0$ and $\ls$ is taken negligible (or for nonzero $\ms$ just replace $\omega \to \sqrt{\omega^2-\ms^2}$). However, this ansatz does not capture the tail at large radius, nor does it account for the fact that $f(0)>0$ is more likely at small $Q$. 

Here, we propose a refined small-$Q$ ansatz $s(r) = s_0 [1-\tanh^2 (\omega' r)]$ and $\Delta f = 1 - f = \pi_0 [1-\tanh^2(\omega' r)]$. 
We must determine $\omega, \omega', s_0, \pi_0$ by eliminating some in favor of other parameters like the parameter $Q$ and minimizing the mass $m_Q$ with respect to the rest. The equations of motion are not analytically tractable for this ansatz, so we will use an expansion in powers of $r$.

For simplicity, we take $\ms=0$ here, but it can easily be added back by modifying $\Omega \equiv \omega/v$. Also define $\Omega' = \omega'/v$ similarly to $\Omega$. First, use $Q = 4 \pi \Omega \int_{0}^{\infty} d \bar{r} \bar{r}^2 s^2$ to obtain $s_0$:
\begin{equation}
s_0 = 3 \sqrt{\frac{Q \, \Omega'^3}{(2\pi^3-12 \pi) \Omega}} \, .
\end{equation}
Next, use the leading order solution to the EOM $\Delta f'' + \frac{2}{\bar{r}} \Delta f' - \frac{1}{2} \frac{\partial V}{\partial{\Delta f}} =0$~\footnote{The final factor of $1/2$ on the left-hand side arises because $\Delta f$ does not have canonical kinetic term normalization.}
near $r=0$ to obtain
\begin{equation}
\pi_0 = \frac{\lc \, s_0^2}{48 \, \Omega'^2} \, .
\end{equation}

Using these results in the formula for the mass
\beqa
m_Q/v = 4\pi \int d \bar{r} \bar{r}^2 \left(\frac{1}{2} s'^2 + f'^2 + \frac{1}{2} \Omega^2 s^2 + \frac{1}{8}\, \lc\,s^2 \,f^2 + \frac{1}{4} \lphi (1 - f^2 )^2  + \frac{1}{4} \ls s^4 +  \frac{1}{2}\ms^2 s^2  \right)\,,
\eeqa
and keeping contributions up to $\mathcal{O}(Q^3)$ leads to
\begin{equation}
\begin{aligned}
m_Q/v \supset & 4 \pi \int_{0}^{\infty} d \bar{r} \bar{r}^2 \left(\frac{1}{2} s'^2 + \Delta f'^2 + \frac{1}{2} \Omega^2 s^2 + \frac{1}{8} \lc s^2 - \frac{1}{4} \lc s^2 \Delta f + \lphi \Delta f^2 \right)
\\
=& \frac{2 \pi Q \Omega'^2}{5(\pi^2-6) \Omega} + \frac{\pi \lc^2 Q^2 \Omega'}{640 (\pi^2-6)^2 \Omega^2} + \frac{Q \Omega}{2} + \frac{\lc Q}{8 \Omega} + \frac{3(15-2\pi^2) \lc^2 Q^2 \Omega'}{320 \pi (\pi^2-6) \Omega^2} + \frac{\lphi \lc^2 Q^2}{512 \pi (\pi^2-6) \Omega^2 \Omega'} \, ,
\end{aligned}
\end{equation}
where the terms in the second line correspond in order to those in the first line. A simple solution for $\Omega'$ can be determined by neglecting the last term and minimizing $m_Q$ with respect to $\Omega'$ for the rest of the terms, leading to 
\begin{equation}
\Omega' = \frac{(11\pi^2 - 90) Q \lc^2}{512 \pi^3 (\pi^2-6) \Omega} \, .
\end{equation}
A slightly more complicated expression for $\Omega$ (not displayed here) results from minimizing the same terms with respect to $\Omega$ after substituting for $\Omega'$. It approaches to its upper limit \cite{Ponton:2019hux} $\Omega < \sqrt{\lc}/2$ as $Q \to 0$ and is relatively independent of $Q$ for the sufficiently small $Q$ valid in our approximations.

The last term in $m_Q/v$ plays a crucial role in setting the minimum charge $\Qmin$. After some algebra, it can be shown that the mass goes as 
\begin{equation}
m_Q/v \sim Q \frac{\sqrt{\lc}}{2} + Q \lphi a_1 - Q^3 a_2 + \mathcal{O}(Q^5) \, ,
\end{equation}
where $a_1$ and $a_2$ are positive constants given in (\ref{eq:M_small}). The second term in this expression comes directly from the $\lphi \Delta f^2$ term in the mass integral, \ie, the vacuum energy from the $\phi$ field. Notice that at sufficiently small $Q$, $m_Q/Q > v \sqrt{\lc}/2 =m_S$. Thus, the Q-ball is unstable at small $Q$. However, at large enough $Q$, the $Q^3$ term dominates the $a_1$ term and makes the Q-ball stable against decaying via evaporation into $Q$ free particles. The value of $\Qmin$ is sensitive in particular to $\lphi$. This approach gives highly accurate predictions for $\Qmin$ compared to the full numerical solutions, as shown for example in Fig.~\ref{fig:MQ-Q}.

\section{Analytic approximations for large $Q$ surface energy}
\label{appedix:surfaceE}

In the large $Q$ limit, it is useful to estimate the surface energy contribution $c_2$ in (\ref{eq:M_large}). (For a simpler derivation of the leading order contributions, see, \eg, \cite{Ponton:2019hux,Bai:2021mzu}.) For a single-field model like Coleman's Q-ball, there is a simple estimation available in the thin-wall limit. If one neglects the friction term $2 s'/\overline{r}$, then the $s$ equation of motion would be 
\begin{equation}
\label{eq:EOM_surface_1field}
    s'' + \frac{\partial U_\text{eff}}{\partial s} = \frac{1}{s'} \left(\frac{1}{2} (s')^2 + U_\text{eff}\right)' = 0 \, ,
\end{equation}
where primes denote derivatives with respect to the radial coordinate $\overline{r}$, and $U_\text{eff} = \Omega^2 s^2/2 - V(s)$. Then, the gradient energy contribution near the surface of the Q-ball can be approximated by
\begin{equation}
\label{eq:M_surface_1field}
    \frac{m_Q}{4 \pi v} \supset \int_0^\infty d\overline{r} \overline{r}^2 \frac{1}{2} s'^2 \approx \frac{1}{2} R^2_Q \int_0^{s_0} ds s' \approx \frac{1}{2} R^2_Q \int_0^{s_0} \sqrt{-2 U_\text{eff}} \, .
\end{equation}
Unfortunately, this approach does not work for multifield potentials like that of (\ref{eq:V}), which contains another degree of freedom $f$ in $U_\text{eff}$. Because the radial field derivative in the middle term of (\ref{eq:EOM_surface_1field}) also picks up a $f' (\partial U_\text{eff} / \partial f)$ term, no simple substitution for $s'$ exists in (\ref{eq:M_surface_1field}).

Thus, to estimate $c_2$ in the model with potential (\ref{eq:V}), we will instead resort to the variation method using the following analytic ansatz for the field solutions in the large-$Q$ limit:
\begin{subequations}
\label{eq:largeQ_surface_ansatz}
\beqa
s &= s_0 \left(1 - \tanh \left(\frac{r-R_Q}{a} \right) \right)  ~, \\
f &= \frac{1}{2} \left(1 + \tanh \left(\frac{r-R_Q}{a} \right) \right) ~.
\eeqa
\end{subequations}
In principle, $R_Q$ and $a$ could be different for the two field profiles to provide a better fit, but keeping them the same simplifies expressions. These provide a reasonable fit to the numerical results. As with any application of the variation method, they will slightly overestimate the true ground state energy of the system.

These field profiles are integrated to obtain expressions for $Q$ and $m_Q$. These integrations will result in polylog functions. We replace the polylogs by their leading-order asymptotic approximations at infinity: $\text{Li}_2(x) \sim -(3 \log^2(x) + \pi^2)/6$ and $\text{Li}_3(x) \sim -(\log^3(x) + \pi^2 \log(x))/6$, checking at each step that these approximations match the full expression to high precision. Then, we substitute the expression for $Q$ to eliminate the variable $s_0$ in the expression for $m_Q$. Leading-order expressions for $\Omega$ and $R_Q$ can be determined by ignoring (surface) terms with $a$ and expanding about $R_Q \to \infty$ at leading order, resulting in the expression $m_{Q\to\infty} \sim Q \Omega / 2 + \pi \lphi R^3_Q /3 + 3 \ls Q^2 / (16 \pi R^3_Q \Omega^2)$. Minimizing with respect to $\Omega$ and $R_Q$ gives $\Omega_{Q\to\infty} = (\lphi \ls)^{1/4}$ and $R_{Q\to\infty} = (3/(4 \pi))^{1/3} \lphi^{-1/4} \ls^{1/12} Q^{1/3}$. \footnote{This agrees with the large-$Q$ result in \cite{Bai:2021mzu} after accounting for differences in the definitions of couplings in (\ref{eq:V}).}

Next, for the surface terms, additional subleading terms in $R_Q$ are added to the expression for $m_Q$ in the previous paragraph: $m_Q = m_{Q\to\infty} + 3 a \ls Q^2 / (64 \pi R^4_Q \Omega^2) + \pi R^2_Q (12 a \lphi + 192 a^{-1})/144$. A new minimized value for $\Omega$ is obtained (including $a$). This new $\Omega$ is substituted into $m_Q$, as well as $R_Q = R_{Q\to\infty} + \Delta R_Q$. The expression is then minimized with respect to $\Delta R_Q$, which in the $Q \to \infty$ limit becomes $\Delta R_Q \sim - 2/(3 a \lphi)$. After further simplification in the $Q \to \infty$ limit and minimizing with respect to $a$, we obtain
\begin{equation}
    c_2 = \left(\frac{\pi}{6}\right)^{1/3} \frac{2 \sqrt{4 \ls^{3/2} \lphi +\ls \lc \sqrt{\lphi}} + \sqrt{4 \lphi^2 \sqrt{\ls} + \lphi^{3/2} \lc}}{2\ls^{1/3} \sqrt{2 \lphi \sqrt{\ls} + \lphi^{3/2}}} \, .
\end{equation}
In actuality, we find that numerically this is only reliable to about a factor of two.

\section{Series expansion near $T_c$}
\label{appendix:bounce_action}

For an FOPT, define a temperature $T_c$ where the two minima of the potential are degenerate, and let $\epsc(T) \equiv (T_c-T)/T_c$. We wish to see the $\epsc$ dependence of relevant quantities in the PT for small $\epsc$. For simplicity, start with a one-field model of the PT with scalar field $\phi$. Then, expanding the potential $V(\phi,T)$ in $\epsc$ for $T \approx T_c$,
\begin{equation}
V(\phi,T) = V(\phi,T_c) - \epsilon_c(T) f(\phi) \, .
\label{eq:Vexp}
\end{equation}
The three-dimensional bounce action for the PT at fixed temperature near $T_c$ is thus (where $\phi=\phi(r)$)
\begin{equation}
\begin{aligned}
S_3 &= 4 \pi \int r^2 dr \left[\frac{1}{2} \left(\frac{\partial\phi}{\partial r}\right)^2 + V(\phi,T) \right]
\\
&= 4 \pi \int r^2 dr \left[\frac{1}{2} \left(\frac{\partial\phi}{\partial r}\right)^2 + V(\phi,T_c) - \epsc f \right]
\\
& \equiv \frac{4 \pi}{2} R^2 S_1 - 4 \pi \epsc \int f\, r^2 dr
\\ 
& \approx \frac{4 \pi}{2} R^2 S_1 - \frac{4 \pi}{3} R^3 \epsc f(\phi_2) \, .
\end{aligned}
\end{equation}
In the last two lines, the thin wall approximation has been assumed---namely, that $f(\phi)\approx f(\phi_2)$ for $r<R$ and $f(\phi) \approx f(\phi_1) \approx 0$ for $r>R$, where $\phi_{1,2}$ are the $T$-dependent minima of $V(\phi,T)$ corresponding to the low-temperature false and true vacua, respectively. The first two terms in the integral in the second line define $S_1$---those terms are only nonzero near the wall at $r \approx R$ and thus contribute a surface term proportional to $R^2$.
Minimizing $S_3$ with respect to $R$, the radius is $R = S_1/(\epsc f(\phi_2))$, and the action is
\begin{equation}
S_3 = \frac{4 \pi\, S_1^3}{6 \,\epsc^2\, f^2(\phi_2)} \, .
\end{equation}

Ultimately, we would like a series expansion of $S_3$ in terms of $\epsilon_c$. We have seen above that we expect the leading term to go as $\epsc^{-2}$ in the thin wall approximation. Thus, we expect the expansion around small $\epsc$ to go as 
\begin{equation}
\label{eq:S3T}
\frac{S_3}{T} = \frac{a}{\epsc^2} + \frac{b}{\epsc} + ...
\end{equation}
To get a sense of how this may look, consider the thermal potential in Ref.~\cite{Dine:1992vs}
\begin{equation}\label{eq:V_DLHLL}
V(\phi, T) = D \,(T^2-T_0^2) \,\phi^2 - E \,T\, \phi^3 + \frac{\lambda}{4}\, \phi^4 \, .
\end{equation}
Here, $T_c = T_0 \sqrt{\lambda D / (\lambda D - E^2)}$, which requires $\lambda D > E^2$. That reference gives a good approximation for the bounce action
\begin{equation}\label{eq:S3_DLHLL}
\frac{S_3}{T} \approx \frac{13.7 D^{3/2} (T^2-T_0^2)^{3/2}}{E^2 T^{3}} \left[1+ \frac{\alpha}{4} \left(1+ \frac{2.4}{1-\alpha} + \frac{0.26}{(1-\alpha)^2}\right) \right] \,
\end{equation}
with $\alpha \equiv \lambda D (T^2-T_0^2)/(E^2 T^2)$. A series expansion of this in powers of $\epsc$ reveals the coefficients $a,b$ as
\begin{align}\label{eq:S3oT_ab}
a = \frac{0.22 E^5}{\lambda^{3/2} (\lambda D - E^2)^2} ~,
\qquad \qquad 
b  = \frac{3.00 E^3 (\lambda D - 1.22 E^2)}{\lambda^{3/2} (\lambda D - E^2)^2} ~.
\end{align}

In the model in \cite{Dine:1992vs}---which considers the electroweak PT---$D \propto g^2$ and $E \propto g^3$, where $g$ is the $SU(2)$ gauge coupling. A similar argument may apply to our model, \eg, if one of the scalar fields is gauged. Thus, for small $g$, $a\propto \lambda^{-7/2} g^{11}$. It can be checked that $(\epsn b)/a \propto g^{3/2}$ in the small $g$ limit, where we have used $\epsn \propto a^{1/2}$ derived in Appendix \ref{appendix:nucleation_sites}. Therefore, the expansion in (\ref{eq:S3T}) can hold to arbitrarily small $g$ and thus arbitrarily small $a$. For example, with $g \sim 10^{-3}$ one may have $a \sim 10^{-30}$.

\section{Number density of nucleation sites}
\label{appendix:nucleation_sites}

With the parametrized bounce action $S_3/T$, it is straightforward to determine the number density of bubble nucleation sites.
We start from the bubble nucleation rate per volume $\gamma$, which is written as
\begin{align}
	\gamma \approx \, T^4 \, \left( \frac{S_3}{2\pi T}\right)^{3/2} \, e^{-S_3/T} \,,
\end{align}
where we have omitted an $\mc{O}(1)$ coefficient.
The fraction of space in the false volume can then be calculated as
\begin{align}
	h(t) = \mathrm{exp}\Bigl[ - \frac{4\pi}{3} \int^t_{t_c} \! dt' \, v_\mathrm{sh}^3\,(t-t')^3\, \gamma(t') \Bigr] \,,
\end{align}
where $t_c$ is the time at which the plasma temperature equals $T_c$, and $v_{\rm sh}$ is the bubble wall velocity. 
As the spacial fraction of the unbroken phase is exponentially suppressed when the PT starts, we take $h(t_n)=1/e$ as the definition of the bubble nucleation time $t_n$ (the corresponding temperature $T_n$ is the bubble nucleation temperature).
We use the saddle-point approximation to evaluate the integration in the exponential function, as it is saturated at late time.
Specifically, we rewrite $\gamma$ as $\gamma(t^\prime) = \mathrm{exp}{[\log \gamma(t^\prime)]}$ and expand $\log\gamma$ to the next-to-leading order as $\log{\gamma(t^\prime)} \approx \log{ \gamma(t_n)} +\, (t^\prime - t_n)\, \xi$.
Defining
\begin{align}
	\beta \equiv - \frac{d(S_3/T)}{dt} = \left( \frac{\dot{T}/T}{-H} \right) \left( T \frac{d(S_3/T)}{dT} \right) H \,,
\end{align}
we express $\xi$ as
\begin{align}
	\xi 
	\equiv \frac{d}{dt} \log \gamma 
	= \beta - \frac{3}{2}\, \frac{\beta}{(S_3 / T) } + 4 \, \frac{\dot{T}}{T}\approx \beta \,.
\end{align}
With these approximations, the bubble nucleation time can be determined as
\begin{align}\label{eq:t_n}
h(t_n)=1/e\ \Rightarrow\ \frac{4\pi}{3} \int_{t_c}^{t_n} \! d t' \, v_\mathrm{sh}^3 \, (t_n - t')^3\, \gamma(t_n) \, e^{(t^\prime - t_n) \beta} \ \approx \ 8\pi v_\mathrm{sh}^3 \gamma(t_n) \, \beta^{-4}\ \approx\ 1\,.
\end{align}
The number density of nucleation site can be correspondingly estimated as
\begin{align}
    n_\mathrm{nuc} 
	= \int_{t_c}^{t_n} \! d t^\prime \, \gamma(t^\prime) \, h(t^\prime) 
	\, \approx \, \bigl( 8 \pi v_\mathrm{sh}^3 \,\beta^{-3} \bigr)^{-1} ~.
\end{align}

With the parametrization 
in (\ref{eq:S3T}),
it is more convenient to perform the calculations in terms of the supercooling parameter $\epsc$ instead of the physical time $t$. 
It is easy to show that $\beta/H\approx T\ d(S_3/T)/dT=(1-\epsc)(2a/\epsc^3+b/\epsc^2)$. 
The supercooling parameter $\epsc$ at the nucleation time can thus be determined from \eqref{eq:t_n} as
\begin{align}
    8\pi\, v^3_{\rm sh}\,T^4_n\left(\frac{a/\epsn^2+b/\epsn}{2\pi}\right)^{3/2}=\exp\left(\frac{a}{\epsn^2}+\frac{b}{\epsn}\right)\left[(1-\epsn)\left(\frac{2a}{\epsn^3}+\frac{b}{\epsn^2}\right)\right]^4 H^4_n\,,
\end{align}
where the subscript $n$ indicates that the corresponding quantity is evaluated at $t=t_n$.
As we expect the exponential factor on the right-hand side to dominate, we can reorganize the equation as
\begin{align}
    \frac{a}{\epsn^2}+\frac{b}{\epsn}\ =\ \log\left(\frac{8\pi\, v^3_{\rm sh}\,T^4_n\left(\frac{a/\epsn^2+b/\epsn}{2\pi}\right)^{3/2}}{\left[(1-\epsn)\left(\frac{2a}{\epsn^3}+\frac{b}{\epsn^2}\right)\right]^4 H^4_n}\right)\ \equiv\ c\,.
\end{align}
We therefore have $\epsn=(2a)/(-b\pm\sqrt{b^2+4ac})$.
Note that there is still a mild $\epsn$ dependence in $c$.

\begin{figure}[t]
	\centering
	\includegraphics[width=0.49\textwidth]{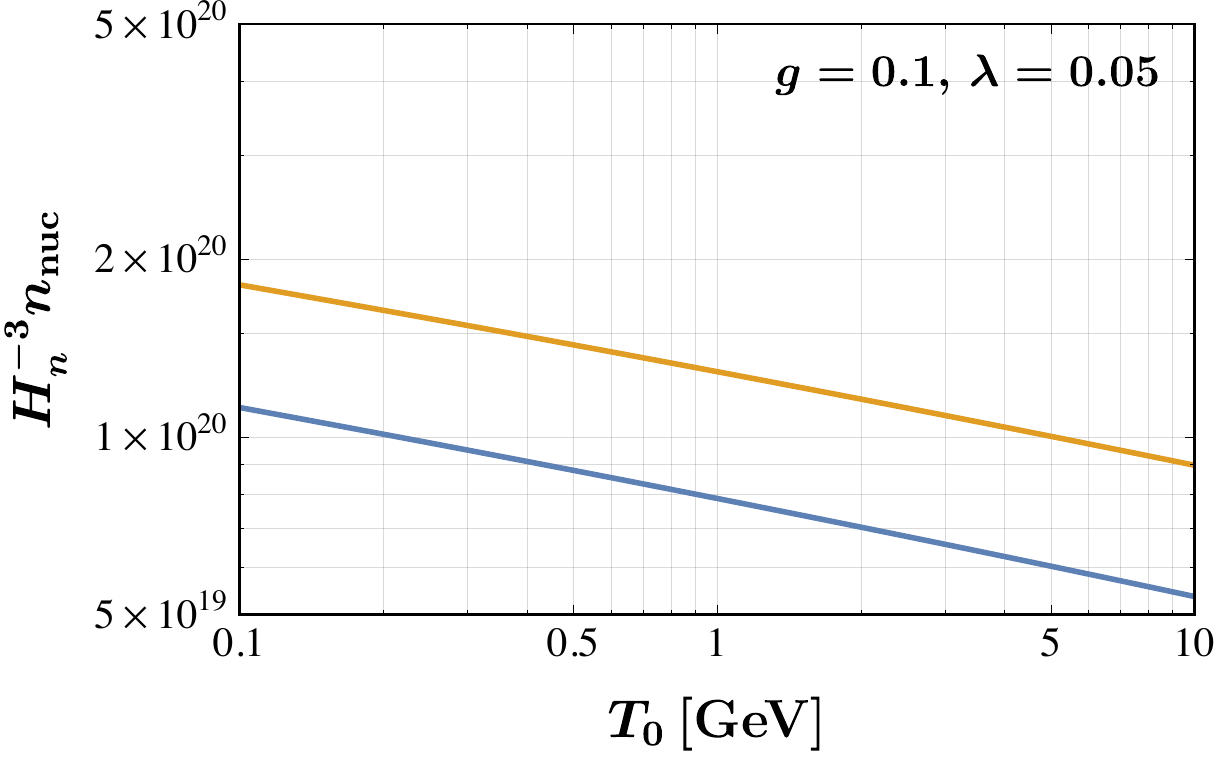}
	\includegraphics[width=0.49\textwidth]{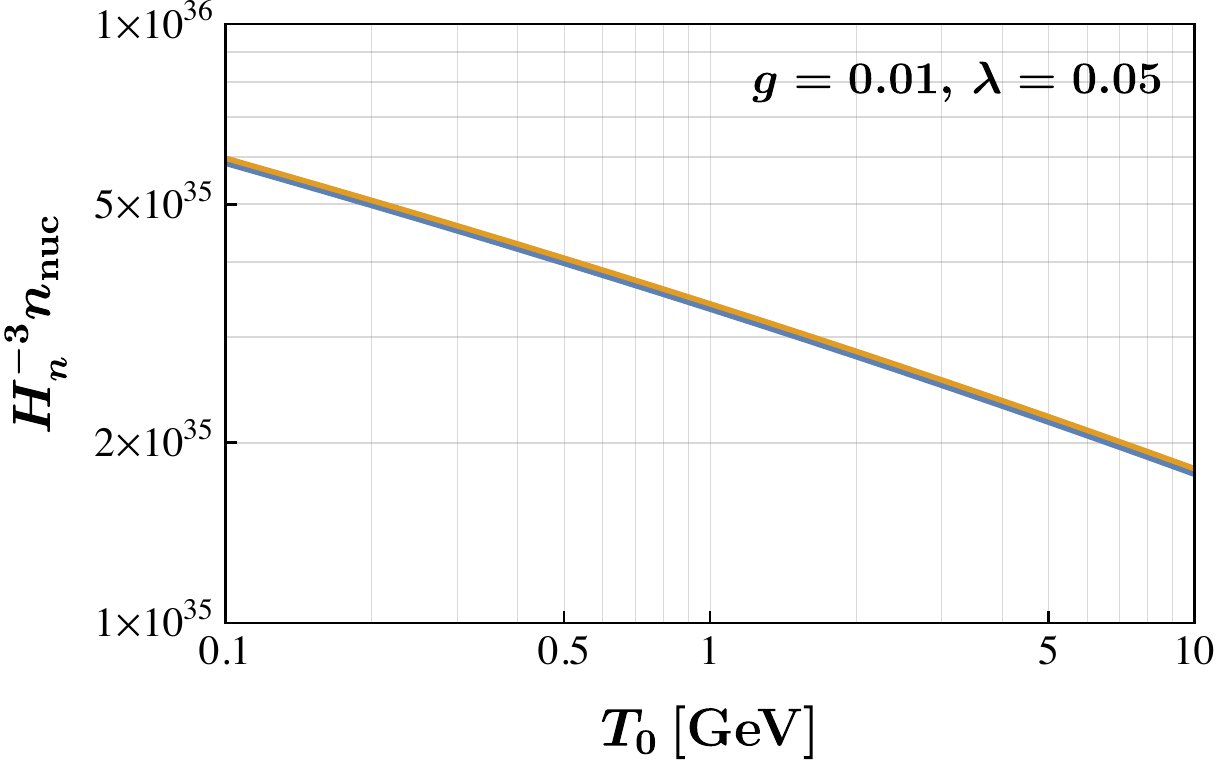}
	\caption{
	The number of nucleation sites in a Hubble patch at the bubble nucleation temperature $T_n$ for the potential in \eqref{eq:V_DLHLL}.
	The blue line includes only the $a$ term in the expansion \eqref{eq:S3T} as in \eqref{eq:S3oT_ab}, while the orange curve includes both the $a$ and $b$ terms in the expansion.
	We have chosen $D=g^2$ and $E=g^3$ as argued at the end of Appendix~\ref{appendix:bounce_action}, and the two panels are plotted with $g=0.1$ and 0.01 respectively.
	Here we approximately have $T_n\approx T_0$ for the parameters considered. 
	}
	\label{fig:S3_NLO}
\end{figure}

Under the thin-wall approximation, Appendix~\ref{appendix:bounce_action} showed that the bounce action satisfies $S_3/T\propto \epsc^{-2}$, {\it i.e.} $b=0$. 
It is worth checking to what extent this approximation is valid, or how much the $b/\epsc$ term contributes to the bounce action $S_3/T$ and the nucleation site number density $n_{\rm nuc}$. 
We consider the potential in \eqref{eq:V_DLHLL} and keep only the $1/\epsc^2$ and $1/\epsc$ terms in the fitted bounce action \eqref{eq:S3_DLHLL}.
The results are shown in Fig.~\ref{fig:S3_NLO}. 
The difference in the number density of bubble nucleation sites brought by the NLO term is in general within one order of magnitude compared with the LO-only result.

For the limit $b=0$, a simpler expression results: $a = \epsn^2 \log [ v_\text{sh}^3 \epsn^9 T_n^4 / (8 \sqrt{2\pi} a^{5/2} H_n^4)]$ (where $\epsn \ll 1$ was assumed). Thus, for the expansion in powers of $\epsn$ to hold, $\epsn \lesssim 1$ implies $a \lesssim 100$ for $T_n \sim 100~\GeV$. Then, $n_\text{nuc} \approx (4 \pi v_\text{sh}^3 a^{1/2})^{-1} H_n^3 (\log [v_\text{sh}^3 \epsn^9 T_n^4 / (8 \sqrt{2\pi} a^{5/2} H_n^4)])^{3/2}$. 
Note that a larger value for $a$ gives a later PT with less bubble nucleation. Thus, larger $a$ is associated with a larger charge $Q$.


\providecommand{\href}[2]{#2}\begingroup\raggedright\endgroup

\end{document}